\documentclass[journal]{IEEEtran}

\usepackage[utf8]{inputenc}
\usepackage[T1]{fontenc}
\usepackage{amsmath,amsfonts}
\usepackage{amssymb}
\usepackage{amsthm}
\usepackage{epsfig}
\usepackage{calc}
\usepackage[dvipsnames]{xcolor}
\usepackage[all]{xy}
\usepackage{bm}
\usepackage{algorithm}
\usepackage{algpseudocode}%
\usepackage{enumerate}
\usepackage{bbm}
\usepackage{url}
\usepackage{enumerate}
\usepackage{framed}
\usepackage{caption}
\usepackage{blkarray}
\usepackage{enumitem}
\usepackage{array}
\usepackage{textcomp}
\usepackage{stfloats}
\usepackage{verbatim}
\usepackage{todonotes}
\usepackage{tikz}
\usepackage{cite}             
\usepackage{graphicx}         
\usepackage{booktabs}         
\usepackage[caption=false,font=normalsize,labelfont=sf,textfont=sf]{subfig}
\newtheorem{defn}{Definition}
\newtheorem{thm}{Theorem}[section]
\newtheorem{cor}[thm]{Corollary}
\newtheorem{prop}[thm]{Proposition}
\newtheorem{pty}{Property}

\newtheorem{lem}[thm]{Lemma}
\newtheorem{conj}[thm]{Conjecture}
\newtheorem{constr}[thm]{Construction}
\newtheorem{note}{Remark}

\newcommand{\bit}{\begin{itemize}}
\newcommand{\eit}{\end{itemize}}
\newcommand{\bcor}{\begin{cor}}
\newcommand{\ecor}{\end{cor}}
\newcommand{\beq}{\begin{equation}}
\newcommand{\eeq}{\end{equation}}
\newcommand{\beqn}{\begin{equation*}}
\newcommand{\eeqn}{\end{equation*}}
\newcommand{\bea}{\begin{eqnarray}}
\newcommand{\eea}{\end{eqnarray}}
\newcommand{\bean}{\begin{eqnarray*}}
\newcommand{\eean}{\end{eqnarray*}}
\newcommand{\ben}{\begin{enumerate}}
\newcommand{\een}{\end{enumerate}}
\newcommand{\bdefn}{\begin{defn}}
\newcommand{\edefn}{\end{defn}}
\newcommand{\bnote}{\begin{note}}
\newcommand{\enote}{\end{note}}
\newcommand{\bprop}{\begin{prop}}
\newcommand{\eprop}{\end{prop}}
\newcommand{\bpty}{\begin{pty}}
\newcommand{\epty}{\end{pty}}
\newcommand{\blem}{\begin{lem}}
\newcommand{\elem}{\end{lem}}
\newcommand{\bthm}{\begin{thm}}
\newcommand{\ethm}{\end{thm}}
\newcommand{\bconj}{\begin{conj}}
\newcommand{\econj}{\end{conj}}
\newcommand{\bconstr}{\begin{constr}}
\newcommand{\econstr}{\end{constr}}
\newcommand{\bpf}{\begin{proof}}
\newcommand{\epf}{\end{proof}}

\newcommand\mtiny[1]{\mbox{\tiny\ensuremath{#1}}}
\newcommand{\revp}[1]{\textcolor{black}{#1}}

\begin{document}

\title{Block Circulant Codes with Application to Decentralized Systems}

\author{Birenjith Sasidharan~\IEEEmembership{Member,~IEEE}, Emanuele Viterbo~\IEEEmembership{Fellow,~IEEE}, Son Hoang Dau~\IEEEmembership{Member,~IEEE}
\thanks{B.\ Sasidharan and E.\ Viterbo are with the Electrical and Computer System Eng. Dept., Monash University, Clayton, Australia (e-mail: \{birenjith.padmakumarisasidharan, emanuele.viterbo\}@monash.edu), S.H. Dau is with 
RMIT University, Melbourne, Australia,
(e-mail: sonhoang.dau@rmit.edu.au)}
\thanks{This work is supported by the Australian Research Council through the Discovery Project under Grant DP200100731.}}

\markboth{Journal of \LaTeX\ Class Files,~Vol.~xx, No.~x, Sxxx~xxxx}%
{Shell \MakeLowercase{\textit{et al.}}: A Sample Article Using IEEEtran.cls for IEEE Journals}

\IEEEpubid{}

\maketitle

\begin{abstract} 
In this paper, we design a family of $[n,k,d]$ block circulant codes that consist of many $[n_0 \ll n,k_0 \ll k,d_0]$ local codes and that satisfy three properties: (1) the code supports distributed erasure decoding, (2) $d$ can be scaled above $d_0$ by a given parameter, and (3) it is amenable to low complexity verification of code symbols using a cryptographic commitment scheme. These properties make the code ideal for use in protocols that address the data availability problem in blockchain networks. Moreover, the code outperforms the currently used 2D Reed-Solomon (RS) code with a larger relative minimum distance $(d/n)$, as desired in the protocol, for a given rate $(k/n)$ in the high-rate regime.

The code is designed in two steps. First, we develop the topology, i.e., the structure of linear dependence relations among code symbols, and define it as the block circulant topology $T_{[\mu,\lambda,\omega]}(\rho)$. In this topology, there are $\mu$ local codes, each constrained by $\rho$ parity checks. The set of symbols of a local code intersects with another in a uniform pattern, determined by two parameters, namely the overlap factor $\lambda$ and the overlap width $\omega$. Next, we instantiate the topology, i.e., to specify the coefficients of linear dependence relations, to construct the block circulant codes ${\cal C}_{\text{BC}}[\mu,\lambda,\omega,\rho]$. Every local code is a $[\lambda\omega+\rho,\lambda\omega,\rho+1]$ generalized RS code. The block circulant code has $n=\mu(\rho+\omega)$, $k=\mu\omega$ and we show that $d=\lambda\rho+1$ under certain conditions. For $\lambda=2$, we prove that $d=2\rho+1$ always, and provide an efficient, parallelizable erasure-correcting decoder that fully recovers the codeword when there are $\leq 2\rho$ erasures. The decoder uses a novel decoding mechanism that iteratively recovers erasures from pairs of local codes.
\end{abstract}
 
\begin{IEEEkeywords}
block circulant code, topology, blockchain, data availability, data availability sampling, decentralization, locality. 
\end{IEEEkeywords}

\section{Introduction\label{sec:intro}}

\IEEEPARstart{I}{n} many applications, the code design problem comes with additional structural constraints, in addition to meeting the targets on the error-correction capability and rate. The linear dependence relations between symbols in a codeword are expected to conform to certain structures, as demanded by the needs of the application. For example, in codes with locality \cite{GopHSY12}, every symbol is part of a set of linear dependence relations involving a small number of symbols, much less than the dimension of the code. In a distributed storage system, this structure helps repair a failed node by downloading data from only a small number of easily accessible live nodes. The locality constraint does not set
the coefficients of the linear dependence relations and specifies only that the number of symbols involved in the relation must be small. In this direction, a new approach emerged in the field of code design \cite{GopHKSWY2017}. Desirable features of a code developed for an application are achieved in two logically separated steps, namely, first by determining the structure of linear dependence relations among code symbols and second by specifying the coefficients involved in these relations. The first is referred to as the {\em topology} of the code, and the second as an {\em instantiation} of the chosen topology.  This two-step process of designing linear codes is referred to as the {\em modular approach} \cite{GopHKSWY2017}. 

The modular approach in code design became prominent along with the development of locally repairable codes\cite{GopHSY12}. The notions of locality  \cite{GopHSY12,HuaCL2013,PraKLK12}, locality with global parities\cite{HuaCL2013}, availability \cite{WanZ14,BalK17}, and hierarchical locality \cite{SasAK15} are examples of diverse topologies that have received attention due to applications in distributed storage. Although it is formally articulated only recently in \cite{GopHKSWY2017}, the modular approach has existed in coding theory since its early days. For instance, the product topology is quite well known starting from its introduction by Elias in 1954 \cite{Elias54} with the aim of reliable communication with reduced decoding complexity, and up to its rejuvenated interest owing to applications in solid-state drives \cite{BlaHH2013,PlaB14}. Many product-like topologies were introduced later. The product topology with additional global constraints binding all code symbols is named the grid-like topology and was explored in \cite{GopHKSWY2017} in an attempt to characterize all correctable erasure patterns. The staircase codes \cite{SmiFHKL2012,SheKS23} and diamond codes \cite{BagT97} with applications, respectively, in fiber optic communication and magnetic disk arrays, are codes with product-like topology with controlled deviations from the product topology, as required by the type of error/erasure patterns to be corrected. \revp{More variants of grid-like topology are explored in literature for different applications\cite{BlaH18,BlaH2020,HouHLWHB2023}}.

This paper adopts the modular approach to design erasure codes that are suitable for a unique application in {\em decentralized systems}. \revp{A decentralized system consists of {\em untrusted nodes} and must have a protocol to verify the legitimacy of data transfer and execution of algorithms. This verification is essential to guarantee {\em safety} (that the system does not settle to a ``bad'' state) and {\em liveness} (that the system eventually settles to a ``good'' state) of the system. Unlike in a typical distributed system, it is challenging in a decentralized system to seamlessly distribute data storage and algorithms' execution across nodes in the network, without sacrificing safety and liveness. This necessitates different forms of centralization of the storage and algorithms, thereby leading to scalability issues. 
}

\revp{
A typical example of a decentralized system is a blockchain network that realizes a written-in-stone ledger of transactions. In its simplest form, a blockchain network relies on resource-rich `full' nodes, each storing the complete transaction history as a chain of blocks. Later, a more scalable solution introduced resource-efficient `light' nodes, which require fewer resources. The light nodes only store block headers (i.e., a ``summary'' of the full block data) 
and are designed to respect safety and liveness. However, this new approach results in a security vulnerability known as the {\em data availability (DA) problem}{\footnote{The data availability problem in a decentralized system\cite{BasSBK2021} is quite different from the problem of availability in a distributed storage system \cite{WanZ14}. \revp{In fact, the entire codeword may even be stored in a single node in the context of DA problem whereas symbols are stored in different nodes in the setting of a distributed storage system.}}} \cite{BasSBK2021}. 
}  

The DA problem arises in a blockchain network when a malicious full node hides a few data symbols, and there is no way to determine if the unavailability of data is due to malicious intentions or to network delays/failures. \revp{This can eventually drive the system to a ``bad'' state where the states of light nodes and full nodes are inconsistent with each other. A robust protocol that tackles the DA problem includes two components: (i) every full node stores data after encoding it using a $[n,k,d]$ linear code, and (ii) every light node independently queries for a tiny number $s \ll n$ of randomly coded symbols. The protocol must meet two conditions: (a) If more than $(d-1)$ symbols of the codeword are hidden/unavailable, then the majority of light nodes must query at least one hidden symbol with high probability to guarantee safety; (b) If more than $(n-d+1)$ symbols are available in full nodes, then light nodes must collectively query at least $(n-d+1)$ symbols to guarantee liveness. In the latter case, the protocol also involves distributed decoding of the unavailable symbols, followed by a broadcast of these reconstructed symbols among the full nodes.} 

\revp{It is wise to choose an $[n,k,d]$ code with both the rate $(k/n)$ and the relative minimum distance $(d/n)$ as large as possible, first to reduce the storage overhead at every full node and second, to increase the probability for a light node to query a hidden symbol. At the same time, the code is expected to exhibit certain properties. A necessary condition for a code to permit distributed decoding is the following:} 
\bit
\item \revp{{\bf Property 1}: [{\em Distributed decoding}] An $[n,k,d]$ code is amenable to distributed decoding of erasures only if it consists of $[n_0 \ll n,k_0 \ll k,d_0]$ local codes (i.e., codes obtained by puncturing certain locations) and the collection of parity check (p-c) constraints of all the local codes spans the space of maximum dimension $(n-k)$. }
\eit 
\revp{Maximum distance separable (MDS) codes such as Reed-Solomon (RS) codes, do not have such property. In locally repairable codes, particularly the optimal constructions \cite{HuaCL2013,TamB2014,PraKLK12,SonDYL14,SilRV2015}, global p-c constraints (i.e., involving all $n$ code symbols) are added to the constraints of $[n_0,k_0,d_0]$ local codes to raise $d$ above $d_0$. Therefore, they also fail to have the above property. Moreover, this calls for a second property:}
\bit
\item \revp{{\bf Property 2}: [{\em Minimum distance scaling without global parity checks}] An $[n,k,d]$ code consisting of local codes of parameters $[n_0,k_0,d_0]$ is said to scale $d$ beyond $d_0$ without global parity checks if its $d$ can be made strictly larger than $d_0$ while satisfying Property 1.}
\eit
\revp{It turns out that a code with product topology exhibits both Properties 1 and 2. When a coded symbol is communicated over the network, a cryptographic tool, known as the {\em commitment scheme} is used to verify that the received symbol is legitimate. It is necessary to check whether the symbol has not been tampered with and whether it is part of a valid codeword. There are polynomial commitment schemes, such as the well-known Kate-Zaverucha-Goldberg (KZG) scheme~\cite{KateZG10} with optimized implementations~\cite{Sup2023}, that allow these checks to be carried out with reduced complexity, provided the local codes are polynomial evaluation codes with block length $n_0 \ll n$. This leads to the third property that relates to instantiation of the desired topology:}
\bit
\item \revp{{\bf Property 3}: [{\em Commitment Compatibility}] An $[n,k,d]$ code is commitment compatible if every $[n_0 \ll n,k_0 \ll k, d_0]$ local code is compatible with a commitment scheme, e.g., the RS code with the KZG scheme.} 
\eit 
\revp{Quite appropriately, the 2D RS code with the largest $(d/n)$ among product codes has been chosen in practice~\cite{BasSBK2021} because it satisfies all three above properties. However, its relative minimum distance $(d/n)$ drops significantly when the rate $(k/n)$ is high. In this paper, we present a new {\em block circulant topology} and a family of {\em block circulant (BC) codes} instantiating the topology that achieves a higher $(d/n)$ than that of product codes in the high rate regime ($\geq 0.5$), while enjoying all the three desired properties to tackle the DA problem. The BC codes we introduce are the only known codes, apart from 2D RS codes, that satisfy all the desired properties.} 



In further detail, we make the following contributions:
\ben 
\item 
We propose a new block circulant topology $T_{[\mu,\lambda,\omega]}(\rho)$ characterized by a collection of $\mu\rho$ p-c constraints associated with $\mu$ local codes each determined by $\rho$ parity checks. The regular pattern of intersection among the supports of local codes is determined by two parameters: the {\em overlap factor} $\lambda$ and the {\em overlap width} $\omega$. The topology exhibits Properties $1$ and $2$ as desired. 
\item 
We construct $[n=\mu(\rho+\omega), k=\mu\omega, d]$ block circulant codes ${\cal C}_{\text{BC}}[\mu,\lambda,\omega,\rho]$ that instantiate the block circulant topology with $[n_0=\lambda\omega+\rho, k_0=\lambda\omega, d_0=\rho+1]$ Reed-Solomon codes as its local codes, thus exhibiting Property 3. Under certain conditions, we derive that $d=\lambda\rho+1$ thus scaling beyond $d_0$ by about a factor of $\lambda$ as required by Property 2. We also provide an efficient distributed decoding algorithm to correct $(d-1)$ erasures when $\lambda=2$. In particular, we propose a novel mechanism for decoding erasures beyond the capability of local codes using \textit{pairs} of consecutive local codes. 
\item
We show that BC codes offer a high-rate alternative to 2D RS codes, thereby achieving a wider parameter regime for a protocol designed to tackle the DA problem. For example, if we compare the performance of a $[1444,1024,49]$ 2D RS code with a $[1416,1024,65]$ shortened BC code both having the same storage overhead of $\tilde 1.4$x, the BC code requires to query only $53$ coded symbols per light node in a system of $1000$ light nodes against $72$ required by the 2D RS code to achieve the same target performance. Apart from that, the lower number of $12$ of local codes in the BC code instead of $76$ in the 2D RS code yields a proportionate reduction in both the space- and time-complexity of realizing the KZG scheme. 
\een 
In Sec.~\ref{sec:bctop}, \revp{we present the new block circulant topology in light of the product topology.} In Sect.~\ref{sec:code2}, we describe a code construction that instantiates the topology for an overlap factor $\lambda=2$, determines its minimum distance, and provides an efficient erasure-correcting decoding algorithm. The construction is generalized for every $\lambda \geq 2$ in Sect.~\ref{sec:codegt2}. Finally, in Sect.~\ref{sec:bc}, we discuss in detail the application of the code in a protocol that addresses the DA problem. 

{\bf Notation:} We define $[a] := \{1,2,\ldots, a\}$ and $[a, b] := \{a, a+1, \ldots, b\}$ for any integers $a, b$. For any vector ${\bf x} = (x_1 \ x_2, \ \ldots, \ x_n)$, $\textsl{supp}({\bf x}) = \{ i \mid 1 \leq i \leq n, x_i \neq 0 \}$. Let $A = \{a_1, a_2, \ldots, a_{|A|}\} \subseteq [n]$. Then we denote the vector $(x_{a_i}, i=1,2,\ldots, |A|)$ of length $|A|$ by ${\bf x}_A$ and sometimes ${\bf x}|_A$. A punctured code ${\cal C}_A = \{{\bf c}_A \mid {\bf c} \in {\cal C}\}$ of an $[n,k,d]$ code ${\cal C}$ is known as a {\em local code} of ${\cal C}$ if $k_0\ll k$ and $n_0\ll n$ when ${\cal C}_A$ is an $[n_0,k_0,d_0]$ code. We call $A$ the support of ${\cal C}_A$. An $(a\times a)$ identity matrix is denoted by $I_a$. Occasionally, we write a matrix $B$ of size $(m\times n)$ as $B_{m \times n}$ to improve readability. A vector of evaluations of a polynomial $p(X)$ at points in an ordered set $S$ is denoted by $p(S)$. 

\section{\revp{The Block Circulant Topology\label{sec:bctop}}}
\begin{figure}
       \captionsetup[subfloat]{justification=centering}
     \centering
     \subfloat[][Square product]{\includegraphics[width=1.28in]{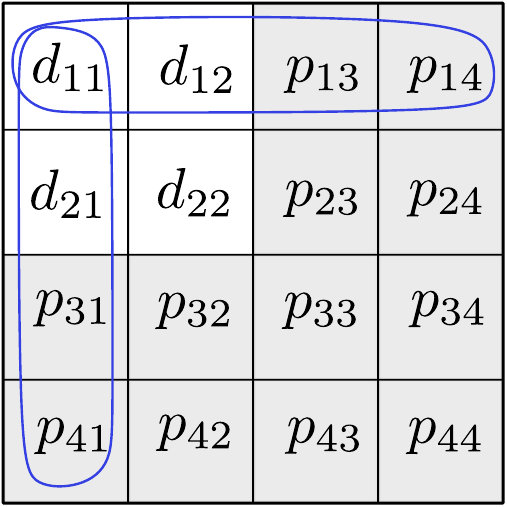}\label{fig:prtop}}
     \subfloat[][Block circulant]{\includegraphics[trim = -10mm 0mm 0mm 0mm,width=1.65in]{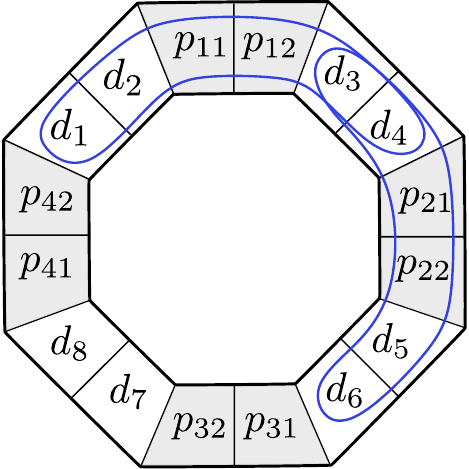}\label{fig:bctop}}
     \caption{\revp{Illustration of the square product and the block circulant topologies. It may be viewed that we start with information symbols of the unshaded region and extend with parity symbols of the shaded region resulting in the final arrangement. The groupings of symbols depicted by closed curves indicate local codes.}}
     \label{fig:top}
\end{figure}
In this section, we first outline the product topology and then introduce the block circulant topology, relating it to the former for a clear comparison. In \cite{GopHKSWY2017}, Gopalan et al. defined a grid-like topology, which is a unified framework for product codes augmented by global parity check constraints. Here, we restrict their definition by removing global parity check constraints that are not needed in the context of this work.

\bdefn \cite{GopHKSWY2017} (Product topology) Let $m \leq \ell$ be positive integers. Consider an $\revp{(m \times \ell)}$ array of symbols $x_{ij}, 1\leq i \leq m, 1 \leq j \leq \ell$ from some field $\mathbb{F}$. Let $0 \leq a \leq m-1$ and $0 \leq b \leq \ell-1$ be integers. Then we define $T_{m\times \ell}(a,b)$ as product topology as follows: 
\ben 
\item There are `$a$' parity check equations for each column $j \in [\ell]$, i.e.,
\beqn
\sum_{i=1}^{m} \alpha_{i}^{(u)} x_{ij} \ = \ 0,~~ u \in [a]
\eeqn 
for \revp{some field elements} $\{ \alpha_{i}^{(u)} \}$.
\item There are `$b$' parity check equations for each row  $i \in [m]$, i.e.,
\beqn
\sum_{j=1}^{\ell} \beta_{j}^{(u)} x_{ij} \ = \ 0, ~~u \in [b]
\eeqn 
for \revp{some field elements} $\{ \beta_{j}^{(u)} \}$.
\een 
A linear code over a field $\mathbb{F}$ is said to instantiate the product topology $T_{m\times \ell}(a,b)$ if it has a parity check matrix $H$ that is completely determined by parity check equations of $T_{m\times \ell}(a,b)$ for some fixed assignment of values to $\{ \alpha_{i}^{(u)} \},  \{ \beta_{j}^{(u)} \}$.

We call the special case $T_{n_0\times n_0}(n_0-k_0,n_0-k_0)$ where $m=\ell=n_0$,  $a=b=n_0-k_0$ for some $k_0 < n_0$ as the square product topology.
\edefn 

A code instantiating the product topology shall have two local codes associated with every symbol, namely an $[m,m-a]$ code, and an $[\ell,\ell-b]$ code, the former referred to as the {\em column code} and the latter as the {\em row code}. The generator matrix of a code instantiating $T_{m\times \ell}(a,b)$ takes the form of tensor product of generator matrices of row and column codes. The square product topology is a specialization obtained when both the row and column codes have the same parameters $[n_0,k_0,d_0]$. For example, when $n_0=4,k_0=2$, the generator matrix $G_{\text{prod}}$ takes the form:
\bea \label{eq:prod} 
G_{\text{prod}} \ = \ \left[ \begin{array}{cccc}
     1 & & g_{\mtiny{13}} & g_{\mtiny{14}} \\
      & 1 & g_{\mtiny{23}} & g_{\mtiny{24}} 
\end{array} \right] \otimes \left[ \begin{array}{cccc}
     1 & & g_{\mtiny{13}} & g_{\mtiny{14}} \\
      & 1 & g_{\mtiny{23}} & g_{\mtiny{24}} 
\end{array} \right] 
\eea 
where $\otimes$ represents the tensor product of matrices. 

As an important example, the 2D Reed-Solomon code ${\cal C}_{\text{2DRS}}$ is an instantiation of the square product topology $T_{n_0\times n_0}(n_0-k_0,n_0-k_0)$ with every local (row/column) code being an $[n_0,k_0,n_0-k_0+1]$ RS code. The ${\cal C}_{\text{2DRS}}$  having parameters $[n=n_0^2,k=k_0^2,d=(n_0-k_0+1)^2]$ comes with the following features, which will be relevant in a decentralized system:
\ben
\item 
There exist distributed decoding and encoding algorithms in which every participant node (except for any coordinating node) needs to run algorithms only related to an $[n_0,k_0]$ RS code.
\item \revp{For any $[n, k, d]$ code instantiating $T_{m\times \ell}(a,b)$, it is known that 
\bea \label{eq:dgrid}
d & \leq & (a+1)(b+1).
\eea
The code ${\cal C}_{\text{2DRS}}$ attains equality in \eqref{eq:dgrid} when $a=b=n_0-k_0$.}
\item Any algorithm to verify the correctness of a code symbol suffices to have access only to the row and column codewords associated with that symbol. 
\een 

\revp{ 
As an example, the square product topology $T_{4\times 4}(2,2)$ is illustrated in Fig.~\ref{fig:prtop},
based on the generator matrix in \eqref{eq:prod}. We begin with a $2\times 2$ grid of $k_0^2=4$ {\em information symbols} that are not bound to any constraint. Every row and column can then be extended with $2$ parity symbols such that the resultant $4 \times 4$ grid of $n_0^2=16$ symbols satisfy all constraints of the square product topology. Hence, each symbol, whether information or parity, becomes part of two local codes. The rate is $k_0^2/n_0^2=0.25$. }

\revp{ 
Next, we introduce the {\em block circulant topology} with a simple example by arranging $16$ symbols in a circle as shown in Fig.~\ref{fig:bctop}. We begin with $k=8$ unconstrained information symbols ($d_1,\ldots, d_8$) placed in $\mu=4$ groups of size  $2$, equally spaced on the circle. We add $\rho=2$ parity symbols $(p_{11},p_{12}),\ldots, (p_{41},p_{42})$ between every two adjacent groups to end up with a total of $n=16$ symbols. Information symbols in every two adjacent groups along with parity symbols in between are subjected to linear constraints to form a local code. Then we shall have $\mu=4$ local codes. Observe that every information symbol is part of $\lambda=2$ local codes, whereas every parity symbol belongs to only one. The supports of two adjacent local codes intersect at $\omega=2$ symbols. We say that the arrangement has an {\em overlap width} of $\omega=2$ and an {\em overlap factor} of $\lambda=2$. Suppose that we index the entire set of symbols in the clockwise order, starting from $d_1$ until $p_{42}$ as $(x_0,x_1,\ldots, x_{15})$. We can view the set of symbols as the union of four sets $X_1=\{x_0,x_1,\ldots,x_5\}, X_2=\{x_4,x_5,\ldots,x_9\}, X_3=\{x_8,x_9,\ldots,x_{13}\}$ and $X_4=\{x_{12},x_{13},\ldots,x_{15}\}$. Here, $X_i,i=1,2,3,4$ contains symbols that are constrained to form a local code. We may denote the supports of local codes as $A_1=[0, 5], A_2=[4,9], A_3=[8,13]$ and finally $A_4=[12,15]$. Then, we can view each of these supports as a union of three pairwise disjoint subsets, for example $A_1 = B_{11} \cup B_{10} \cup B_{12}$ where $B_{11}=\{0,1\}, B_{10}=\{2,3\},$ and $B_{12}=\{4,5\}$. Here, $B_{11}$ corresponds to a subset of information symbols that intersects with the local code on the left, and $B_{12}$ to the one intersecting with the local code on the right. The set $B_{10}$ corresponds to the set of parity symbols.  Every local code is a $[\lambda\omega+\rho=6,\lambda\omega=2]$ code. Overall, we obtain a $[16,8]$ code, and observe that the rate $(k/n)=0.5$ increases two-fold with respect to the square product topology. This arrangement is formally generalized in the following definition.}

\begin{figure*}
\bea \label{eq:bctoph}
H_{\text{BC-top}} & \hspace{-2mm} = & \hspace{-3mm}\left[\begin{array}{cccccccccccc}
     H_{11} & H_{12} & H_{13} &  & H_{15} & & & & & & & \\
     & & H_{23} & H_{24} & H_{25} & & H_{27} 
    & & & & &  \\
     & & & & H_{35} & H_{36} & H_{37} & & H_{39} & & & \\
     & & & & & & H_{47} & H_{48} & H_{49} & & H_{4,11} &  \\
     H_{51} & & & & & & & & H_{59} & H_{5,10} & H_{5,11} & \\
     H_{61} & & H_{63} & & & & & & & & H_{6,11} & H_{6,12}
\end{array} \right]_{\mu\rho \times \mu(\rho+\omega)}, \\
H_{ij}  & \hspace{-2mm}= & \hspace{-3mm}\left\{ \hspace{-2mm}\begin{array}{ll}
    \left[ \begin{array}{cc}
     h_{2i-1,(5j/2)-1} & h_{2i-1,(5j/2)} \\
     h_{2i,(5j/2)-1} & h_{2i,(5j/2)}
\end{array} \right]_{\rho \times \rho}, & \hspace{-2mm}  i \in [6], \ j=2i,  \\
\left[ \begin{array}{ccc}
     h_{2i-1, ((5j-1)/2)-1 } & h_{2i-1,(5j-1)/2} & h_{2i-1,((5j-1)/2)+1} \\
     h_{2i,((5j-1)/2)-1} & h_{2i,(5j-1)/2} & h_{2i,((5j-1)/2)+1}
\end{array} \right]_{\rho \times \omega} \hspace{-3mm}, &\hspace{-2mm}  i \in [6], \ j= 2i-1,2i+1,2i+3 \!\!\! \mod 12~.
\end{array} \right. 
\eea
	\caption{The structure of parity check matrix in block circulant topology $T_{[\mu,\lambda,\omega]}(\rho)$ when $\mu=6, \lambda=3,\omega=3,\rho=2$.}
\end{figure*}

\bdefn \label{def:bctop} (Block circulant topology) Let $\lambda \geq 2, \omega,\rho$ be positive integers. Let $\mu = \lambda\nu$ be a multiple of $\lambda$ by an integer $\nu \geq 1$. Consider a set of  $n = \mu(\rho + \omega)$ symbols $\{x_0, x_1, \ldots, x_{n-1}\}$ from some field $\mathbb{F}$. Let $A_i, i=1,2,\ldots, \mu$ be (intersecting) subsets of $[0, n-1]$ each of size $\rho + \lambda\omega$.
We define $A_i$ as the union of $(\lambda+1)$ pairwise disjoint subsets $B_{i0},B_{i1}, \ldots, B_{i,\lambda}$ 
\begin{equation*}
A_i = \bigcup_{j=0}^{\lambda} B_{ij}, ~~\ i \in [\mu],
\end{equation*}
where $B_{ij}, i =1,2,\ldots, \mu, j =0,1,\ldots, \lambda$ are given by  
\bean
B_{ij} &\!\!\!\! = &\!\!\!\! \{(i+j-2)(\rho +\omega ),(i+j-2)(\rho +\omega )+1, \ldots, \\&& \hspace{1mm}
(i+j-2)(\rho +\omega )+\omega-1 \}, ~~\ j \in [\lambda], \  i\in [\mu] ,\\
B_{i0} &\!\!\!\! = &\!\!\!\! \{ i\omega+(i-1)\rho , i\omega+(i-1)\rho +1 , \ldots, \\&& \hspace{35mm} 
i(\rho +\omega )-1 \}, ~~\ i \in [\mu], 
\eean 
and every element in $B_{ij}, j=0,1,\ldots, \lambda$ is computed modulo $\mu(\rho+\omega)$. 
\revp{Note that $|B_{i0}|=\rho$ and $|B_{ij}|=\omega$ when $j>0$, and $|A_i|=\lambda \omega+\rho$.}
The set of symbols $\{x_0, x_1, \ldots, x_{n-1}\}$ is identified as a union of $\mu$ subsets 
\bean
X_i & = & \{x_j \mid j \in A_i\}, \ i \in [\mu],
\eean 
each of size $\lambda \omega + \rho$.  
We index the elements of $X_i$ alternatively as $X_i = \{x_{i0}, x_{i1},\ldots, x_{i,\lambda\omega+\rho-1}\}$ for every $i \in [\mu]$. Then we define $T_{[\mu,\lambda,\omega]}(\rho)$ as block circulant topology which constitutes the following: There are $\rho$ parity check equations for every $i \in [\mu]$, i.e., 
\beqn
\sum_{j=0}^{\lambda\omega+\rho-1} \alpha_{ij}^{(u)} x_{ij} \ = \ 0, \ u \in [\rho], \ i \in [\mu]
\eeqn 
for \revp{some field elements} $\{ \alpha_{ij}^{(u)} \}$.

A linear code over a field $\mathbb{F}$ is said to instantiate the block circulant topology $T_{[\mu,\lambda,\omega]}(\rho)$ if it has a parity check matrix $H$ that is completely determined by parity check equations of $T_{[\mu,\lambda,\omega]}(\rho)$ for some fixed assignment of values from $\mathbb{F}$ to $\{ \alpha_{ij}^{(u)} \}$.
\edefn 
The structure of parity check matrix of a code instantiating $T_{[\mu,\lambda,\omega]}(\rho)$ is given in \eqref{eq:bctoph} taking an example with $\mu = 6, \lambda = 3, \omega = 3, \rho = 2$. \revp{If we imagine every $H_{ij}$ as a single entity, then each block row of $H_{\text{BC-top}}$ is cyclically shifted by two blocks relative to the previous one. This structure inspires the topology's name.} It can be observed that the submatrix of $H_{\text{BC-top}}$ formed by columns associated to $\{H_{ij}\mid j \text{ is even}\}$ can be made full-rank by appropriate choice of coefficients. In that case, all the rows of the $H_{\text{BC-top}}$ become linearly independent, making it a full-rank matrix. We shall restrict our discussion to only such codes instantiating $T_{[\mu,\lambda,\omega]}(\rho)$.

\revp{A prominent feature of $H_{\text{BC-top}}$ is the regular pattern of intersection between the supports of local codes. We say {\em a local code overlaps with another} if the intersection of their supports is non-empty. If $\lambda < \mu$ then every local code overlaps with $2(\lambda - 1)$ other local codes, each on a subset of $\omega$ symbols. If $\lambda=\mu$, every local code overlaps with $(\lambda-1)$ other local codes. For this reason, we call $\lambda$ the {\em overlap factor} and $\omega$  the {\em overlap width} in $T_{[\mu,\lambda,\omega]}(\rho)$. In the example shown in \eqref{eq:bctoph}, we have $\lambda=3 < \mu$, and every local code overlaps with $4$ other local codes. } 

\revp{Consider an $[n,k,d]$ code ${\cal C}$ instantiating $T_{[\mu,\lambda,\omega]}(\rho)$. It is clear that blocklength $n=\mu(\rho+\omega)$ and dimension $k=\mu \omega$. The code satisfies Property 1 as constraints of local codes fully determine the topology. All the local codes in ${\cal C}$ have the same length $n_0=\lambda\omega+\rho$, dimension $k_0=\lambda\omega$  and rate of $R_{0}=\frac{k_0}{n_0}=\frac{\lambda\omega}{\rho+\lambda\omega}$. The code ${\cal C}$ has a rate 
\bea 
R  & = &\frac{k}{n}= \frac{\omega}{\rho+\omega}. 
\eea 
Although $R < R_0$, it is possible to achieve arbitrarily high rates by adjusting $(\rho/\omega)$. } 

\revp{While the minimum distance of every local code is bounded by $d_0 \leq n_0-k_0+1=\rho + 1$, it is not obvious how large a minimum distance can the overall code achieve. In the coming two sections, we provide a code construction ${\cal C}_{\text{BC}}[\mu,\lambda,\omega,\rho]$ instantiating $T_{[\mu,\lambda,\omega]}(\rho)$, that achieve a minimum distance of $\lambda\rho + 1$. This establishes that $T_{[\mu,\lambda,\omega]}(\rho)$ allows to scale $d$ beyond $d_0$ by a factor of about $\lambda$, thus satisfying Property 2. The construction uses polynomial evaluation codes as its local codes, and hence satisfies Property 3 as well as described in Sec.~\ref{sec:intro}.} 

\section{Construction of Block Circulant Codes for Overlap Factor $\lambda=2$\label{sec:code2}}

In this section, we present the construction of a code  ${\cal C}_{\text{BC}}[\mu,2,\omega,\rho]$ that instantiates $T_{[\mu,\lambda,\omega]}(\rho)$ with $\lambda=2$. We begin with a review of generalized Reed-Solomon (GRS) codes as they form an important building block for the construction.

\subsection{A Review of Generalized Reed-Solomon Codes\label{sec:grs}}

Let $\mathbb{F}_q$ be a finite field of $q > n$ elements. Let $\Lambda = \{ \alpha_0,\alpha_1,\ldots, \alpha_{n-1}\}$ be a set of $n$ non-zero elements in $\mathbb{F}_q$. Let
\bea 
V_r(\Lambda) & := & \left[\begin{array}{cccc}
	1 & 1 & \cdots & 1 \\
	\alpha_0 & \alpha_1  & \cdots & \alpha_{n-1}  \\
	\vdots & \vdots & & \vdots \\
	\alpha_0^{r-1} & \alpha_1^{r-1}  & \cdots & \alpha_{n-1}^{r-1}
\end{array} \right] 
\eea
denote an $(r \times n)$ Vandermonde matrix.  When it is required to make explicit reference to elements, we may as well write $V_r(\Lambda)$ as $V_r(\alpha_0,\alpha_1,\ldots, \alpha_{n-1})$. For some non-zero elements ${\beta}_0,{\beta}_1,\ldots,{\beta}_{n-1} \in \mathbb{F}_q$, we construct an $(n\times n)$ diagonal matrix $M$ and define a parity check matrix $H_{\text{GRS}}$ as follows:
\bea
M & = & \textsl{diag}({\beta}_0,{\beta}_1,\ldots,{\beta}_{n-1}) \\
H_{\text{GRS}} & = & V_{r}(\Lambda) \cdot M  .
\eea
A GRS code over $\mathbb{F}_q$ is defined as
\bea\label{eq:cgrs1}
{\cal C}_{\text{GRS}} = \{{\bf c} \in \mathbb{F}_q^n \ | \ H_{\text{GRS}}{\bf c}^T = {\bf 0} \} .
\eea 
It is well known that ${\cal C}_{\text{GRS}}$ is an $[n,k=n-r,d=r+1]$ linear code and can alternatively be defined as an evaluation code:
\bea \label{eq:cgrs2}
{\cal C}_{\text{GRS}}   = && \hspace{-5mm}\{{\bf c} = ({\beta}'_0f(\alpha_0), {\beta}'_1f(\alpha_1),\ldots, {\beta}'_{n-1}f(\alpha_{n-1})) \ | \nonumber \\
&&\ f(X) \in  \mathbb{F}_q[X], \ \textsl{deg}(f(X)) \leq k-1 \}
\eea 
where ${\beta}'_0,{\beta}'_1,\ldots,{\beta}'_{n-1} \in \mathbb{F}_q$ are constants given by
\bea \label{eq:a}
({\beta}'_i)^{-1} & = & {\beta}_i \prod_{\substack{j=0 \\ j \neq i}}^{n-1} (\alpha_i - \alpha_j), \ i = 0,1,\ldots,n-1 .
\eea 
We refer to $\alpha_j, {\beta}_j, {\beta}'_j$ respectively as {\em  evaluation point} or {\em code locator}, {\em parity check column multiplier}, and {\em generator column multiplier} all associated to the code symbol $c_j, j=0,1,\ldots, n-1$ of the codeword ${\bf c}=(c_0,c_1,\ldots, c_{n-1})$. The polynomial $f(X)$ that defines a codeword ${\bf c} \in {\cal C}_{\text{GRS}}$ as given in \eqref{eq:cgrs2} is referred to as the {\em message polynomial} of ${\bf c}$. From \eqref{eq:a}, it follows that we can fix ${\beta}_i'=1$ for all $i$ by suitably choosing ${\beta}_i, i=0,1,\ldots,n-1$ as a function of code locators. We refer to such a GRS code as the {\em basic Reed-Solomon} (BRS) code defined as 
\bea \label{eq:brs}
{\cal C}_{\text{BRS}} &\!\!  = &\!\!  \{{\bf c} =(f(\alpha_0), f(\alpha_1),\ldots, f(\alpha_{n-1})) \ |  \\  && ~~~f(X) \in \mathbb{F}_q[X],
\ \text{deg}(f(X)) \leq k-1 \}.\nonumber
\eea 
A set of code locators completely determines a BRS code. Every codeword ${\bf c} \in {\cal C}_{\text{BRS}}$ is mapped to a message polynomial $f(X)$ by an $1$-$1$ correspondence. We denote this bijection by $\Phi_{{\cal C}_{\text{BRS}}}$ and write $\Phi_{{\cal C}_{\text{BRS}}}(f) = {\bf c}$. When the code is clear from the context, we may omit the subscript and write $\Phi(f) = {\bf c}$. 

\begin{figure*}
	\beq 
	\label{eq:hbc2} H_{\text{BC},2} \ = \ \left[ \begin{array}{ccccccccccccc}
		W_{11} & W_{12} & W_{13} & & & & & & & & && \\
		& & W_{23} & W_{24} & W_{21} & & & &  & & &&\\
		& & & & W_{11} & W_{12} & W_{13} & & & & &&\\
        & & & & &  & W_{23} & W_{24} & W_{21} & & &&\\
		& &  &  &  &  & & & \ddots & \ddots &  &&\\
		    && & &  &  &  &  & & W_{11} & W_{12} & W_{13}&  \\
		W_{21} & & && & & & & & & & W_{23} & W_{24}
	\end{array} \right]_{2\nu\rho \times 2\nu(\rho+\omega)}.
\eeq 
	\caption{The parity check marix of ${\cal C}_{\text{BC}}[\mu,2,\omega,\rho]$ that instantiates  $T_{[\mu,2,\omega]}(\rho)$. We have $\mu=\lambda\nu=2\nu$. The submatrix  $W_{ij}=V_jM_{ij}$ where $V_j$'s are Vandermonde matrices and $M_{ij}$'s are diagonal matrices.}
\end{figure*}

\subsection{The Construction of ${\cal C}_{\text{BC}}[\mu,2,\omega,\rho]$ \label{sec:con2}} 

\revp{The code construction involves specifying coefficients of a parity check matrix whose structure is already prescribed by Definition~\ref{def:bctop}. Let $\mu, \lambda, \omega, \rho$ be integers as defined in $T_{[\mu,\lambda,\omega]}(\rho)$ and let us fix $\lambda=2$, then we have $\mu = 2\nu$ for $\nu \geq 1$. The first two out of $\mu$ block rows of the parity check matrix are designed to correspond to two distinct $[2\omega+\rho,2\omega,\rho+1]$ GRS local codes. The remaining $(\mu-2)$ block rows are obtained by repeating the first pair of block rows $(\mu/2)$ times.
The GRS codes are designed so that a code symbol belonging to two consecutive local codes has the same evaluation point in both. We describe the code construction in a step-by-step manner as follows.} 
\ben 
\item Let $\mathbb{F}_q$ be such that $q > 2(\rho+\omega)$. We pick a subset of non-zero elements $\Lambda = \{ \alpha_0, \alpha_1, \ldots, \alpha_{2(\rho+\omega)-1}\}$ of $\mathbb{F}_q$. Recalling Definition~\ref{def:bctop} for the sets $\{A_i\}, \{B_{ij}\}$ in $T_{[\mu,\lambda,\omega]}(\rho)$, we partition $\Lambda$ into four ordered sets as:
\begin{equation}
\Lambda_{\ell}  :=  \left\{ \begin{array}{ll}
\{\alpha_j \ \mid \ j \in B_{(\ell+1)/2,1} \},& \ \ell = 1, 3 \\
\{\alpha_j \ \mid \ j \in B_{(\ell/2),0} \}, & \ \ell = 2,4 
\end{array} \right. .
\end{equation} 
Observe that $|\Lambda_{\ell}|= \omega$ when $\ell$ is odd and $|\Lambda_{\ell}|= \rho$ when $\ell$ is even. 
\item In association with each $\Lambda_{\ell}$, we define a Vandermonde matrix (by slight abuse of notation) as:
\bea
V_{\ell} & \triangleq & V_{\rho} (\Lambda_{\ell}), \ \ell = 1,2,3,4 .
\eea 
\item Construct diagonal matrices $M_{11}, M_{12}, M_{13}, M_{23}$, $M_{24}$, and $M_{21}$ in such a manner that the matrices $W_1, W_2$ defined below
\bea 
&&\hspace{-10mm}W_1  {=} [W_{11} \ W_{12} \ W_{13}] \ {\triangleq} \ [V_1 M_{11} \ V_2 M_{12} \ V_3 M_{13}] \\
&&\hspace{-10mm}W_2  {=}   [W_{23} \ W_{24} \ W_{21}] \ {\triangleq} \ [V_3 M_{23} \ V_4 M_{24} \ V_1 M_{21}]
\eea
form parity check matrices of two $[2\omega+\rho, 2\omega, \rho+1]$ basic RS codes. Subsequently, we define two GRS codes
\bea
{\cal D}_{[2,\omega,\rho]}^{(1)} & = & \{ {\bf c} \in \mathbb{F}_q^{2\omega+\rho} \mid W_1 {\bf c}^T = {\bf 0} \} \\
{\cal D}_{[2,\omega,\rho]}^{(2)} & =  & \{ {\bf c} \in \mathbb{F}_q^{2\omega+\rho} \mid W_2 {\bf c}^T = {\bf 0} \} 
\eea 
that turn out as the local codes of ${\cal C}_{\text{BC}}[\mu,2,\omega,\rho]$, as will be evident soon. From the discussion in Sec.~\ref{sec:grs}, it may be observed that $M_{11}, M_{12}, M_{13}$ are determined by $\Lambda_1 \cup \Lambda_2 \cup \Lambda_3$ whereas $M_{23}, M_{24}, M_{21}$ by $\Lambda_3 \cup \Lambda_4 \cup \Lambda_1$. Observe that the index $i$ indicates the local code and $j$ indicates the associated Vandermonde matrix in $M_{ij}$. 
\item Next, we construct a $(\mu\rho \times \mu(\rho+\omega))$ parity check matrix $H_{\text{BC},2}$ as defined by \eqref{eq:hbc2}. 
\item Finally, we define the block circulant (BC) code:
\bea \label{eq:cbc2}
{\cal C}_{\text{BC}}[\mu,2,\omega,\rho] & = & \{{\bf c}\in \mathbb{F}_q^n \mid H_{\text{BC},2} {\bf c}^T = {\bf 0} \}. 
\eea 
\een 
In $H_{\text{BC},2}$, the submatrix formed by columns associated to $\{ W_{ij}, \text{$j$ is even}\}$ is of full rank. Therefore, the dimension of ${\cal C}_{\text{BC}}[\mu,2,\omega,\rho]$ is equal to $k=\mu\omega$ and thus ${\cal C}_{\text{BC}}[\mu,2,\omega,\rho]$ is a $[n=\mu(\rho+\omega), k=\mu\omega]$ code instantiating $T_{[\mu,2,\omega]}(\rho)$. There are $\mu$ local codes in ${\cal C}_{\text{BC}}[\mu,2,\omega,\rho]$ and we denote them as
\bean
{\cal C}_{\text{BC},i}[2,\omega,\rho] & = & \{ {\bf c}|_{A_i} \ \mid \ {\bf c} \in {\cal C}_{\text{BC}}[\mu,2,\omega,\rho] \}, \ i \in [\mu]. 
\eean 
When $\mu,\lambda,\rho$ are clear from the context, we may respectively write ${\cal C}_{\text{BC}}[2]$ and ${\cal C}_{\text{BC},i}[2]$ in place of ${\cal C}_{\text{BC}}[\mu,2,\omega,\rho]$ and ${\cal C}_{\text{BC},i}[2,\omega,\rho]$. 
Observe that each local code is either ${\cal D}_{[2,\omega,\rho]}^{(1)}$ or ${\cal D}_{[2,\omega,\rho]}^{(2)}$. 

In the coming subsections, we describe a systematic encoding procedure and an efficient erasure-correction decoder for ${\cal C}_{\text{BC}}[2]$. We also characterize the minimum distance of ${\cal C}_{\text{BC}}[2]$. 

\subsection{Systematic Encoding of ${\cal C}_{\text{BC}}[\mu,2,\omega,\rho]$\label{sec:enc2}} 

The square block-diagonal submatrix  of $H_{\text{BC},2}$
\bean
\tilde{H}_{\text{BC},2} & = & \textsl{diag}[ W_{12},W_{24}, \ldots,W_{12},W_{24} ]
\eean 
is invertible because both $W_{12}$ and $W_{24}$ are invertible matrices. Multiplying $H_{\text{BC},2}$ on the left by  $\tilde{H}_{\text{BC},2}^{-1}$ followed by necessary column permutations, we convert $H_{\text{BC},2}$ into its systematic form. In turn, we obtain a systematic generator matrix $G_{\text{BC},2}$ (upto column permutations) for ${\cal C}_{\text{BC}}[\mu,2,\omega,\rho]$ given by:
\begin{align}
\label{eq:syscbc2} G_{\text{BC},2} &= [I_{\mu\omega} \mid -P_{\text{BC},2}], \ \text{ where } \\
P_{\text{BC},2}&\!=\!\! 
 \left[\!\!\begin{array}{cccccc}
	(W_{12}^{-1}W_{11})^T & & \cdots & (W_{24}^{-1}W_{21})^T \\
	(W_{12}^{-1}W_{13})^T & (W_{24}^{-1}W_{23})^T & &  \\
	& (W_{24}^{-1}W_{21})^T &  &  \vdots \\
	 & & \ddots &   \\ 
	 & & &  (W_{24}^{-1}W_{23})^T \end{array} 
\!\!\!\right].
\end{align}
\revp{We remark that each non-zero block of $P_{\text{BC},2}$ is of the form $A(U^{-1}V)^TB$ where both $U,V$ are Vandermonde matrices and $A,B$ are diagonal matrices, and hence is a Cauchy-like matrix.} 
\subsection{An Erasure Correcting Decoder for ${\cal C}_{\text{BC}}[\mu,2,\omega,\rho]$ \label{sec:dec2}}

\revp{In this subsection, we present an iterative decoder that corrects erasures introduced into a codeword in two phases. In the first phase, erasures are corrected relying only on local codes. In the next phase, erasures that cannot be locally decoded will be recovered by collectively decoding pairs of consecutive local codes. Depending on the pattern, there may still be uncorrected erasures after the two phases. In that case, the decoder outputs a vector with the symbols corrected until then and a message that the output vector has uncorrectable erasures. We provide the decoder in Algorithm~\ref{alg:dec} and give an example to illustrate its operation. We also prove a theorem that the decoder invariably corrects every erasure if the number of erasures in the input is $\leq 2\rho$. }

\revp{We set up a few notations and tools before we proceed to describe the decoder. For every integer $a$, we define $(a)_{\mu} = [(a-1) \mod \mu]+1$.} We use the letter $\textsc{e}$ to denote an erasure. For any vector ${\bf x}=(x_i, i \in A)$ indexed by a finite ordered set $A$, we define 
\begin{equation*}
S_E({\bf x}) := \{ i \mid x_i = \textsc{e}\} \ \subseteq \ A, \ \ 
n_E({\bf x}) := |S_E({\bf x})| \ \leq \ |A|. 
\end{equation*}
For an $[n,k,d {=} n-k+1]$ GRS code ${\cal C}_{\text{GRS}}$ over any field, there is an efficient decoder of time complexity $O(n\log^2 n\log\log n)$ \cite{LinTH16} that can recover from any set of $(d-1)$ erasures. 
Let us denote the {\em erasure correcting decoder} of ${\cal C}_{\text{GRS}}$ by $\textsc{DEC}_{{\cal C}_{\text{GRS}}}$. The algorithm $\textsc{DEC}_{{\cal C}_{\text{GRS}}}$ takes as input a vector ${\bf y}$ that is obtained by subjecting a codeword ${\bf x} \in {\cal C}_{\text{GRS}}$ to erasures. It outputs the correct codeword ${\bf x}$ if $n_E({\bf y}) \leq n-k$, or else if $n_E({\bf y}) > n-k$, it outputs ${\bf y}$ itself. We consider ${\cal C}_{\text{GRS}}$ up to column permutations and therefore $\textsc{DEC}_{{\cal C}_{\text{GRS}}}$ works even if the input vector is permuted with a known permutation.  


Next, we define an auxiliary $[n=2(\omega+\rho), k= 2\omega, d=2\rho+1]$ basic RS code ${\cal D}_{[2,\omega,\rho]}^{(1 \cup 2)}$ \revp{whose set of evaluations points is the union of evaluation points of two consecutive local codes:} 
\begin{align}
\nonumber {\cal D}_{[2,\omega,\rho]}^{(1 \cup 2)} & = \{(f(\alpha), \alpha \in \cup_{\ell=1}^{4}\Lambda_{\ell}) \mid f(X) \in \mathbb{F}_q[X],\\
    \label{eq:jointcode2} & \hspace{32mm} \textsl{deg}(f(X)) \leq 2\omega-1 \}    .
\end{align} 
We use $\textsc{DEC}_{{\cal D}^{(1)}[2]}$, $\textsc{DEC}_{{\cal D}^{(2)}[2]}$ and $\textsc{DEC}_{{\cal D}^{(1\cup 2)}[2]}$ respectively to denote decoders of ${\cal D}_{[2,\omega,\rho]}^{(1)}$, ${\cal D}_{[2,\omega,\rho]}^{(2)}$ and ${\cal D}_{[2,\omega,\rho]}^{(1 \cup 2)}$. We make use of them as building blocks to construct the erasure-correcting decoder $\textsc{DEC}_{{\cal C}_{\text{BC}}[2]}$ for ${\cal C}_{\text{BC}}[\mu,2,\omega,\rho]$, as given in Algorithm~\ref{alg:dec}. The Alg.~\ref{alg:dec} takes as input ${\bf r} = (r_j,j=0,1,\ldots, {n-1}) \in (\mathbb{F}_q \cup \{\textsc{e}\})^n$ defined as  
\bean
r_j = \left\{ \begin{array}{ll}
     c_j & \text{if $j$-th location is not erased}  \\
     \textsc{e} &  \text{if $j$-th location is erased}
\end{array} \right. .
\eean 
where ${\bf c} = (c_j, j=1,2,\ldots, {n-1}) \in {\cal C}_{\text{BC}}[\mu,2,\omega,\rho]$. We assume $n_E({\bf r}) > 0$ because there is no need to invoke the decoder otherwise. We use ${\bf \hat{r}}$ to denote the output (decoded codeword) and ${\bf \hat{r}}[t]$ its version in round $t$ of Phase~1. We also use ${\bf \hat{r}}[{t}]\mid_{A_i}$ for the restriction of ${\bf \hat{r}}[t]$ to the set $A_i$, $i\in [\mu]$. Furthermore, let $J_1$ (\textsl{Line} 4) denote the set of indices of local codes that have at most $\rho$ erasures and hence can be locally decoded in Phase~1. Let $J_2$ (\textsl{Lines} 16, 21) denote the set of index \textit{pairs} of (cyclically) consecutive local codes that cannot be locally decoded but can be \textit{collectively} decoded in Phase 2 (having at most $2\rho$ unrecovered erasures). The following example illustrates how the decoder works. 

\begingroup
\begin{algorithm}
	\caption{$\textsc{Dec}_{{\cal C}_{\text{BC}}[2]}$  \newline {\bf Input}: ${\bf r}$ such that $n_E({\bf r}) > 0$ \newline {\bf Output}: ${\bf \hat{r}}=(\hat{r}_j,j=0,1,\ldots,n-1)$ \label{alg:dec}}
    \begin{algorithmic}[1]
    \State Initialize $t \xleftarrow{} 0$; ${\bf \hat{r}}[t] \xleftarrow{} {\bf r}$; 
    \State ${\bf \hat{r}}_{i}[t] \xleftarrow{} {\bf \hat{r}}[{t}]\mid_{A_i}, i \in [\mu]$
    \State Index entries in vectors as ${\bf \hat{r}}[{t}] = (\hat{r}_{j}[t], \ j=0,1,\ldots, n-1)$ and ${\bf \hat{r}}_{i}[t] = ( \hat{r}_{ij}[t], \  j \in A_i)$
    \State $J_1 \ \xleftarrow{} \ \{ i \in [\mu] \mid \revp{0 <} n_E({\bf \hat{r}}_{i}[t]) \leq \rho \}$
    \While{$J_1 \neq \phi$}
        \For{$i \in J_1$}
            \State 
            ${\bf \hat{r}}_{i}[t+1] \xleftarrow{} \left\{ \begin{array}{ll}
            \textsc{DEC}_{{\cal D}^{(1)}[2]}({\bf \hat{r}}_{i}[t]) & \text{ if $j$ is odd} \\
            \textsc{DEC}_{{\cal D}^{(2)}[2]}({\bf \hat{r}}_{i}[t]) & \text{ if $j$ is even}
            \end{array} \right. $
        \EndFor     
        \State 
                $\hat{r}_{j}[t{+}1] \!\xleftarrow{} \!\left\{\hspace{-2mm} \begin{array}{ll}
                \hat{r}_{ij}[t{+}1] &\hspace{-1mm} \text{if } \exists i \in J_1 \text{ and } \hat{r}_{ij}[t+1] \!\neq\! \textsc{e}   \\
                \hat{r}_{j}[t] & \text{otherwise} 
                \end{array} \right.$ 
        \State $J_1 \ \xleftarrow{} \ \{ i \mid \revp{0 <} n_E({\bf \hat{r}}_{i}[t+1]) \leq \rho \}$
        \State ${\bf \hat{r}}_{i}[t+1] \xleftarrow{} {\bf \hat{r}}[t+1] \mid_{A_i}, i \in [\mu]$; $t \xleftarrow{} t+1$
    \EndWhile
    \State ${\bf \hat{r}} \xleftarrow{} {\bf \hat{r}}[t]$; ${\bf \hat{r}}_{i} \xleftarrow{} {\bf \hat{r}}\mid_{A_i}, i \in [\mu]$; $J_2 \xleftarrow{} \phi$ 
    \If{ $\mu=2$ and $n_E({\bf \hat{r}}) > 0$}
        \If{$n_E({\bf \hat{r}}) \leq 2\rho$ }
            \State $J_2 \xleftarrow{} \{ (1,2) \}$
        \Else
            \State Output ``${\bf r}$ has uncorrectable erasures.''
        \EndIf
    \ElsIf{ $\mu\geq 4$ and $n_E({\bf \hat{r}}) > 0$}
        \State 
            $J_2 \xleftarrow{} \left\{ (i,i') \in [\mu]^2 \left| \begin{array}{l}
                i'=(i+1)_{\mu},\\
                0 < n_E({\bf \hat{r}}|_{A_{i} \cup A_{i'}}) \leq 2\rho, \\
                n_E({\bf \hat{r}}|_{A_{(i-1)_{\mu}} \cap A_{i}}) = 0,  \\
                n_E({\bf \hat{r}}|_{A_{i'} \cap A_{(i'+1)_{\mu}}}) = 0.
                \end{array} \right. \right\} $   
        \If{$\{i \in [\mu] \mid n_E({\bf \hat{r}}|_{A_i}) > 0\} \nsubseteq \cup_{(i,i')\in J_2}\{i,i'\}$}        
            \State Output ``${\bf r}$ has uncorrectable erasures.'' 
        \EndIf
    \EndIf 
    \For{$(i_1, i_2) \in J_2$}
        \State Initialize ${\bf \hat{s}} \xleftarrow{} {\bf 0} \in (\mathbb{F}_q \cup \{\textsc{e}\})^{2\omega+\rho}$
        \State Index ${\bf \hat{s}}$ as ${\bf \hat{s}} = (\hat{s}_j, \ j \in A_{i_1})$
        \If{ $\mu \geq 4$}
            \State 
                    $\hat{s}_j  \xleftarrow{}  \left\{ \begin{array}{ll}
                    \hat{r}_j - \hat{r}_{2(\rho+\omega)+j \!\!\! \mod n} & j \in B_{i_1,1} \\
                    \textsc{e} & j \in B_{i_1,0}\\
                    0 & j \in  B_{i_1,2}  
                    \end{array} \right. $
            \State \label{eq:dec4-shat2}
                    ${\bf \hat{s}} \xleftarrow{} \left\{ \begin{array}{ll}
                    \textsc{DEC}_{{\cal D}^{(1)}[2]}({\bf \hat{s}}) & \text{ if $i_1$ is odd} \\
                    \textsc{DEC}_{{\cal D}^{(2)}[2]}({\bf \hat{s}}) & \text{ if $i_1$ is even}
                    \end{array} \right. $
        \EndIf 
        \State $\hat{s}(X) \xleftarrow{} \Phi^{-1}({\bf \hat{s}})$  
        \State Initialize and index ${\bf \hat{u}}_1= (\hat{u}_{1j}, \ j \in A_{i_1} \cup B_{i_2,0})$, and ${\bf \hat{u}}_2 = (\hat{u}_{2j}, \ j \in B_{i_1,0} \cup A_{i_2}) \in (\mathbb{F}_q \cup \{\textsc{e}\})^{2(\rho + \omega)}$

        \State $\hat{u}_{1j} \xleftarrow{}  \left\{ \begin{array}{ll}
                \hat{r}_{j} & j \in A_{i_1} \\
                \hat{r}_{j} + \hat{s}(\alpha_{j\!\!\mod\!2(\rho+\omega)}) & j \in B_{i_2,0}, \hat{r}_{j} \neq \textsc{e}  \\
                \textsc{e} & j \in B_{i_2,0}, \hat{r}_{j} = \textsc{e} 
                \end{array} \right.$
                
        \State $\hat{u}_{2j} \xleftarrow{} \left\{ \begin{array}{ll}
                \hat{r}_{j} & j \in A_{i_2} \\
                \hat{r}_{j} - \hat{s}(\alpha_{j\!\!\mod\!2(\rho+\omega)}) & j \in B_{i_1,0}, \hat{r}_{j} \neq \textsc{e} \\
                \textsc{e} & j \in B_{i_1,0}, \hat{r}_{j} = \textsc{e}
                \end{array} \right.$

        \State 
                ${\bf \hat{u}}_1  \xleftarrow{} \textsc{DEC}_{{\cal D}^{(1   \cup 2)}[2]}({\bf \hat{u}}_1)$
        \State ${\bf \hat{u}}_2  \xleftarrow{} \textsc{DEC}_{{\cal D}^{(1 \cup 2)}[2]}({\bf \hat{u}}_2)$ 
        \State 
            $\hat{r}_{j} \xleftarrow{} \left\{ \begin{array}{ll}
                \hat{u}_{\ell j} & (\ell=1,j \in A_{i_1}) \text{ or } (\ell=2, j \in A_{i_2})   \\
                \hat{r}_{j} & j \notin A_{i_1} \cup A_{i_2}  
                \end{array} \right.$
    \EndFor
    \end{algorithmic}
    \end{algorithm}
\endgroup

\subsubsection*{Example 1 (Decoding algorithm $\textsc{DEC}_{{\cal C}_{\text{BC}}[2]}$)} 
\begin{figure}
\begin{minipage}{3.6in}
     \centering
     \subfloat[][{\footnotesize Erasure pattern}]{\includegraphics[trim = 0mm -6mm -5mm 0mm,width=1.5in]{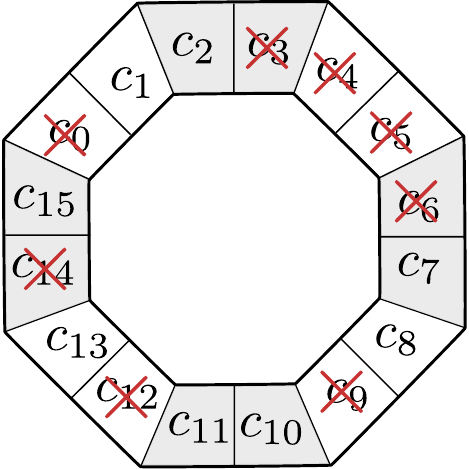}\label{fig:dec1}}
     \subfloat[][{\footnotesize Iteratively decode local codewords with $\leq 2$ erasures.}]{\includegraphics[width=1.64in]{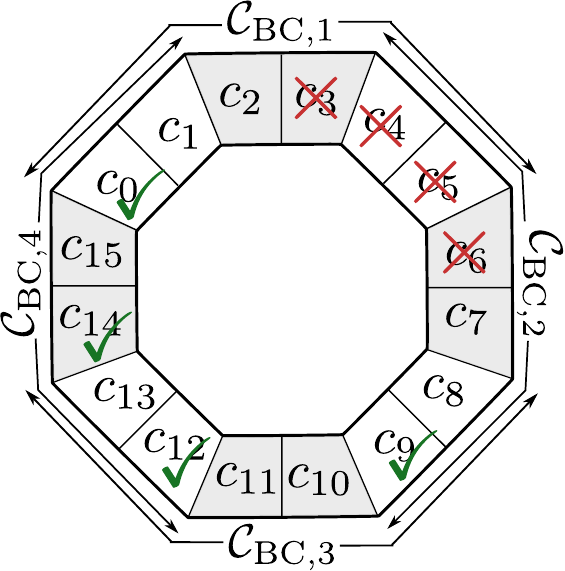}\label{fig:dec2}}
\end{minipage}
\vspace{2mm}
\begin{minipage}{3.6in}
     \centering
     \subfloat[][{\footnotesize Identify code symbols as evaluations of $m_1(X),m_2(X) \in \mathbb{F}_{11}[X]$ at $8$ distinct elements $\alpha_i, i=0,1,\ldots,7$ of $\mathbb{F}_{11}$.}]{\includegraphics[width=3.3in]{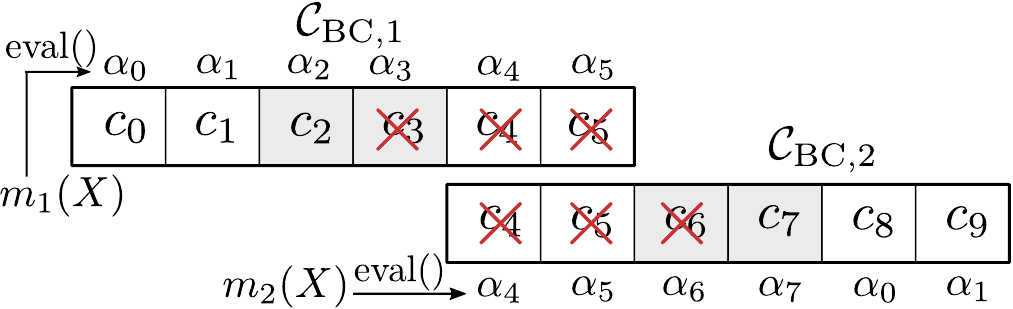}\label{fig:dec3}}
\end{minipage}
\vspace{3mm}
\begin{minipage}{3.6in}
     \centering
     \subfloat[][{\footnotesize Decode $m_1(X)-m_2(X)$ in ${\cal C}_{\text{BC},1}$ and evaluate it at $\{\alpha_6,\alpha_7\}$.}]{\includegraphics[width=3.3in]{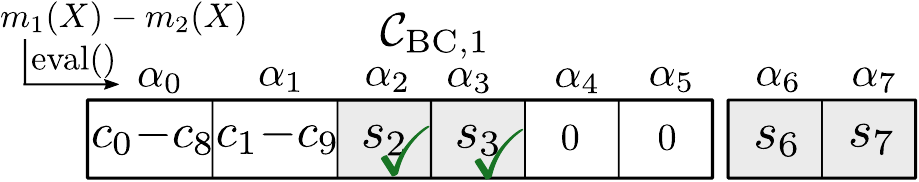}\label{fig:dec4}}
\end{minipage}
\vspace{3mm}
\begin{minipage}{3.6in}
     \centering
     \subfloat[][{\footnotesize Decode $m_i(X),i=1,2$ in ${\cal D}_{{[}2,2,2{]}}^{(1 \cup 2)}$.}]{\includegraphics[width=3.3in]{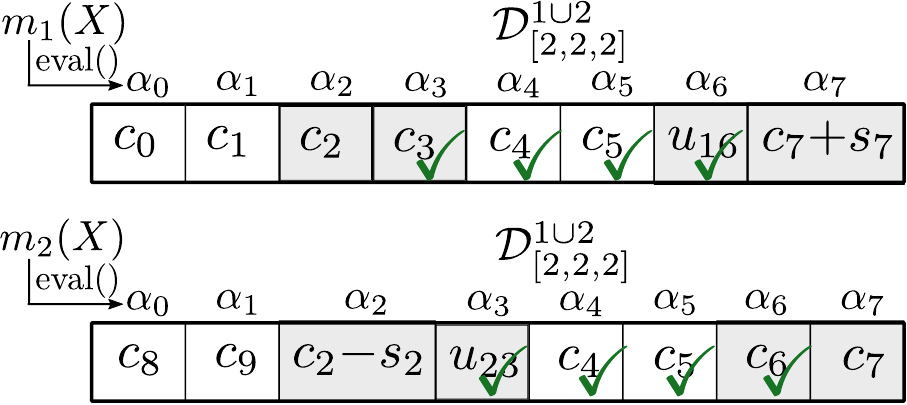}\label{fig:dec5}}
\end{minipage}
\caption{\revp{Illustration of Example 1. We decode a pattern of $8$ erasures in the $[16,8]$ BC code ${\cal C}_{\text{BC}}[4,2,2,2]$ over $\mathbb{F}_{11}$. The BC code has four $[6,4,3]$ GRS local codes ${\cal C}_{\text{BC},i}, i=1,2,3,4$, out of which ${\cal C}_{\text{BC},i}={\cal D}^{(1)}_{{[}2,2,2,{]}}$ for $i=1,3$ and ${\cal C}_{\text{BC},i}={\cal D}^{(2)}_{{[}2,2,2,{]}}$ for $i=2,4$. A symbol with $\times$ indicates an erasure, whereas with $\checkmark$, a correct recovery from erasure. The auxiliary code ${\cal D}_{{[}2,2,2{]}}^{(1 \cup 2)}$ is an  $[8,4,5]$ GRS code defined in \eqref{eq:jointcode2}.  }}
\label{fig:decoder}
\end{figure}

\revp{Consider a codeword ${\bf c} = (c_0,c_1,\ldots, c_{15})$ from the $[n=16,k=8]$ code ${\cal C}_{\text{BC}}[4,2,2,2]$ over the field $\mathbb{F}_{11}$. It is subjected to $8$ erasures at $c_i,i=0,3,4,5,6,9,12,14$. The codeword with erased locations is shown in Fig.~\ref{fig:dec1} assuming the topology of Fig.~\ref{fig:top}a. The code consists of four $[n_0=6,k_0=4,d_0=3]$ local codes ${\cal C}_{\text{BC},i}[2],i=1,2,3,4$, out of which ${\cal C}_{\text{BC},i}[2],i=1,3$ are ${\cal D}_{[2,2,2]}^{(1)}$ (with $\{\alpha_0,\alpha_1,\alpha_2,\alpha_3,\alpha_4,\alpha_5\}$ as evaluation points) and the remaining two are ${\cal D}_{[2,2,2]}^{(2)}$ (with $\{\alpha_4,\alpha_5,\alpha_6,\alpha_7,\alpha_0,\alpha_1\}$ as evaluation points) as marked in Fig.~\ref{fig:dec2}. Note that $\lambda = 2$, $\omega = 2$, $\rho = 2$, and $\mu=4$.}

\revp{In the first phase (\textsl{Lines} $5$-$12$), we decode as many erasures as possible relying only on the local codes. First, $c_9$ and $c_{12}$ are recovered by ${\cal D}_{[2,2,2]}^{(1)}$ and then, $c_{14}$ and $c_0$ by ${\cal D}_{[2,2,2]}^{(2)}$ (see Fig.~\ref{fig:dec2}). Since every local code can correct only up to $d_0-1=2$ erasures, the remaining $4$ erasures at $\{3,4,5,6\}$ can not be corrected by ${\cal D}_{[2,2,2]}^{(1)}$ with support $A_1=[0,5]$, nor by ${\cal D}_{[2,2,2]}^{(2)}$ with support $A_2=[4,9]$.} 

\revp{However, the set of erasure locations $\{3,4,5,6\}$ satisfies the condition checked while computing $J_2$ in \textsl{Line} $21$, i.e., it is $\leq 2\rho$ in size, and it lies exclusively within $A_1 \cup A_2$ without intersecting with $A_3$ or $A_4$. 
Therefore, in the second phase (\textsl{Lines} $26$-$40$), the decoder proceeds to 
correct erasures at $\{3,4,5,6\}$ by decoding the two local codes collectively}.

\revp{Since ${\cal D}_{[2,2,2]}^{(1)}$ and ${\cal D}_{[2,2,2]}^{(2)}$ are $[6,4,3]$ basic RS codes, there are message polynomials $m_1(X)$ and $m_2(X)$ both of degree at most $k_0-1=3$ such that 
\bea \label{eq:eg1c1}
&&\hspace{-12mm}(c_0,c_1,c_2,c_3,c_4,c_5) = m_1(\{\alpha_0,\alpha_1,\alpha_2,\alpha_3,\alpha_4,\alpha_5\}) \\
\label{eq:eg1c2} 
&&\hspace{-12mm}(c_4,c_5,c_6,c_7,c_8,c_9) = m_2(\{\alpha_4,\alpha_5,\alpha_6,\alpha_7,\alpha_0,\alpha_1\}) 
\eea 
where $\alpha_j,j=0,1,\ldots,7$ are $8$ distinct non-zeros elements of $\mathbb{F}_{11}$ (e.g. $\alpha_j = \alpha^j$ for some primitive element $\alpha \in  \mathbb{F}_{11}$). 
Next, consider a vector ${\bf \hat{s}}=(c_0-c_8,c_1-c_9,\textsc{e},\textsc{e},0,0)$. It follows from \eqref{eq:eg1c1} and \eqref{eq:eg1c2} that ${\bf \hat{s}}$ can be obtained by replacing $s_2,s_3$ of a codeword ${\bf s}= (s_0,s_1,s_2,s_3,s_4,s_5) \in  {\cal D}_{[2,2,2]}^{(1)}$ with erasures. It must also be that $s(X)=m_1(X)-m_2(X)$ is the message polynomial of ${\bf s}$ and thus ${\bf s}=(s(\alpha_j),j=0,1,\ldots,5)$. Since there are $k_0=2\omega=4$ unerased symbols in ${\bf \hat{s}}$, it can be correctly decoded to obtain ${\bf s}$, and hence $s(X)$. Evaluations of $s(X)$ at $\alpha_6, \alpha_7$, i.e., $s_6 := s(\alpha_6)$ and $s_7 := s(\alpha_7)$, may as well be computed (see Fig.~\ref{fig:dec3}). Subsequently, we compute 
\bean
m_1(\alpha_7) & = & m_2(\alpha_7)+s(\alpha_7)=c_7+s_7  \\
m_2(\alpha_2) & = & m_1(\alpha_2)-s(\alpha_2)=c_2-s_2,
\eean 
and generate two vectors 
\bean
{\bf \hat{u}}_1 & = &(c_0,c_1,c_2,\textsc{e},\textsc{e},\textsc{e},\textsc{e},c_7+s_7) \\
{\bf \hat{u}}_2 & = & (c_8,c_9,c_2-s_2,\textsc{e},\textsc{e},\textsc{e},\textsc{e},c_7) .
\eean} 
\revp{It is straightforward to check that ${\bf \hat{u}}_1$ and ${\bf \hat{u}}_2$ are associated to two codewords ${\bf u}_1, {\bf u}_2$ of the auxiliary code ${\cal D}_{[2,2,2]}^{(1 \cup 2)}$, respectively with message polynomials $m_1(X)$ and $m_2(X)$. Since ${\cal D}_{[2,2,2]}^{(1 \cup 2)}$ is a $[8,4,5]$ code with evaluation points $\{\alpha_0,\alpha_1,\ldots,\alpha_7\}$, both ${\bf u}_1$ and ${\bf u}_2$ can be correctly decoded from ${\bf \hat{u}}_1, {\bf \hat{u}}_2$. Since the set of evaluation points of the auxiliary code ${\cal D}_{[2,2,2]}^{(1 \cup 2)}$ is precisely the union of evaluation points of ${\cal D}_{[2,2,2]}^{(1)}$ and ${\cal D}_{[2,2,2]}^{(2)}$, decoding of ${\bf u}_1$ and ${\bf u}_2$ yields $m_1(\alpha_3)=c_3$, $m_1(\alpha_4)=m_2(\alpha_4)=c_4$, $m_1(\alpha_5)=m_2(\alpha_5)=c_5$ and $m_2(\alpha_6)=c_6$ (see Fig.~\ref{fig:dec5}). Thus, the codeword ${\bf c}$ is fully recovered.} 
\revp{In the above example, the decoder corrects $(n-k)$ erasures, which is information-theoretically the maximum number of erasures beyond which full recovery is impossible. This is feasible, however, only for certain patterns of erasures. Nevertheless, it can be shown that all erasure patterns containing $\leq 2\rho$ erasures can be correctly decoded by $\textsc{DEC}_{{\cal C}_{\text{BC}}[2]}$. The following theorem proves this assertion, along with establishing the correctness of the algorithm.}

\bthm \label{thm:correct2} Let ${\bf c} =(c_0,c_1,\ldots, c_{n-1}) \in {\cal C}_{\text{BC}}[\mu,2,\omega,\rho]$ and ${\bf r} =(r_0,r_1,\ldots, r_{n-1}) \in (\mathbb{F}_q \cup \{ \textsc{e} \})^n$ be a vector obtained by replacing certain entries of ${\bf c}$ with $\textsc{e}$. Suppose that $\textsc{DEC}_{{\cal C}_{\text{BC}}[2]}$ (Algorithm~\ref{alg:dec}) is invoked with ${\bf r}$ as its input. Then the following statements are true.
\ben
\item[(1)] The algorithm returns ${\bf \hat{r}} =(\hat{r}_0,\hat{r}_1,\ldots,\hat{r}_{n-1}) \in (\mathbb{F}_q \cup \{ \textsc{e} \})^n$ such that  for every $j=0,1,\ldots,n-1$, if $r_j=c_j$ then $\hat{r}_j =c_j$ and if $r_j=\textsc{e}$ then $\hat{r}_j \in \{c_j,\textsc{e}\}$.
\item[(2)] If $n_E({\bf r}) \leq 2\rho$, then the algorithm correctly returns ${\bf c}$. 
\een 
\ethm 
\bpf \revp{The proof is relegated to Appendix~\ref{app:a}. }
\epf 

\subsection{Distributed Decoding of Erasures} 

If decoding of erased symbols in a codeword of an $[n,k,d]$ code $C$ is carried out by a network of nodes with every node (except possibly for a coordinating node that collects the decoded symbols) executing an algorithm that accesses $n_{A} \ll n$ unerased symbols, then we call it a {\em distributed decoding algorithm}. For example, a square product code can be erasure-decoded in a distributed manner with every node accessing unerased symbols associated with a row/column code. \revp{The Property 1 laid out in Sec.~\ref{sec:intro} is a necessary condition for $C$ to permit distributed decoding of erasures. It turns out to be sufficient for ${\cal C}_{\text{BC}}[\mu,2,\omega,\rho]$ because  Alg.~\ref{alg:dec} can be converted to a distributed algorithm as follows. The decoding of erasures happens in two phases, i.e., \textsl{Lines} $5$-$12$ as the first phase, and \textsl{Lines} $26$-$40$ as the second. The remaining lines do not involve decoding any erased symbol and hence can be executed by a coordinating node. In the first phase, the algorithm accesses symbols from only a single local code at a time, i.e., $n_A \leq 2\omega+\rho \ll n$. In the second phase, the algorithm accesses symbols from at most two adjacent local codes at a time, i.e., $n_A \leq 3\omega+2\rho \ll n$. By allocating the execution of each iteration in the first and second phases to nodes in a network working in parallel, distributed decoding can be realized.} As explained in Sec.~\ref{sec:intro}, this is relevant in the application of the code in a decentralized system. \revp{Similar to the case in 2D RS code, a node taking part in distributed decoding of the BC code has to decode a local GRS code of shorter length, say $n_L$. So, the worst-case per-node time-complexity is $O(n_L\log^2 n_L\log\log n_L)$ where $n_L=2(\rho+\omega)=(2n/\mu)$, as opposed to $n_L=\sqrt{n}$ in the 2DRS code.} We remark here that even systematic encoding of the BC code (given in Sec.~\ref{sec:enc2}) can be carried out as a distributed algorithm, quite similar to how product codes support distributed encoding.

\subsection{The Minimum Distance of ${\cal C}_{\text{BC}}[\mu,2,\omega,\rho]$} 
We characterize the minimum distance of the code ${\cal C}_{\text{BC}}[\mu,2,\omega,\rho]$ in the following theorem. 
\bthm \label{thm:dmin2} Let ${\cal C}_{\text{BC}}[\mu,2,\omega,\rho]$ as defined in Sec.~\ref{sec:con2} be an $[n=\mu(\rho+\omega), k =\mu\omega, d]$ linear code. Then 
\bea \label{eq:dmin2}
d & = & 2\rho +1.
\eea 
\ethm
\bpf 
Since ${\cal C}_{\text{BC}}[\mu,2,\omega,\rho]$ is a linear code, 
\bean
d = \min_{\substack{{\bf c} \in {\cal C}_{\text{BC}}[\mu,2,\omega,\rho], \\ {\bf c} 
 \neq 0} } w_H({\bf c})  .
 \eean
Let ${\bf v}^T = [1,0,0,\ldots,0]$  be a $(1\times \mu\omega)$ vector containing all zeros except for a $1$ at the first location. The codeword ${\bf c} \in {\cal C}_{\text{BC}}[\mu,2,\omega,\rho]$ obtained by encoding ${\bf v}$ using the systematic generator matrix $G_{\text{BC},2}$ in \eqref{eq:syscbc2} is the first row of $G_{\text{BC},2}$. Every non-zero block of $G_{\text{BC},2}$  apart from $I_\omega$ is a Cauchy-like matrix and hence all its entries are non-zero. Therefore $w_H({\bf c}) = 2\rho+1$ showing that $d \leq 2\rho+1$. By Theorem~\ref{thm:correct2}, any erasure pattern of at most $2\rho$ erasures can be corrected and therefore $d \geq 2\rho+1$. This completes the proof. 
\epf 

\section{Construction for Every Overlap Factor\label{sec:codegt2}}
In this section, we generalize the construction described in Sec.~\ref{sec:code2} for every $\lambda \geq 2$ and show that the code has a minimum distance of $\lambda\rho+1$ under certain conditions.


Let $\lambda,\omega,\rho, \mu=\lambda\nu$ be positive integers as defined in $T_{[\mu,\lambda,\omega]}(\rho)$. We describe the code construction in a step-by-step manner as follows. 
\ben 
\item Let $\mathbb{F}_q$ be such that $q > \lambda(\rho+\omega)$. We pick a subset of non-zero elements $\Lambda = \{ \alpha_0, \alpha_1, \ldots, \alpha_{\lambda(\rho+\omega)-1}\}$ of $\mathbb{F}_q$. Recall the definition of sets $A_i, B_{ij}$ from Definition~\ref{def:bctop}. We partition $\Lambda$ into $2\lambda$ ordered sets as given below:
\bea
\Lambda_{\ell} &\hspace{-3mm} = &\hspace{-3mm} \left\{ \begin{array}{ll}
\hspace{-2mm} \{\alpha_j \mid   j \in B_{(\ell+1)/2,1} \},&\hspace{-2mm}  \ell {=} 1, 3,\ldots, 2\lambda-1 \\
\hspace{-2mm} \{\alpha_j   \mid j \in B_{(\ell/2),0} \}, &\hspace{-2mm}  \ell {=} 2,4,\ldots, 2\lambda
\end{array} \right. 
\eea 
Observe that $|\Lambda_{\ell}|= \omega$ when $\ell$ is odd and $|\Lambda_{\ell}|= \rho$ when $\ell$ is even. 
\item In association to each $\Lambda_{\ell}$, we define a Vandermonde matrix (by slight abuse of notation) as:
\bea
V_{\ell} & \triangleq & V_{\rho} (\Lambda_{\ell}), ~~~\ell = 1,2,\ldots, 2\lambda . 
\eea 

\item For $i = 1, 2,\ldots, \lambda$, we are interested in constructing $W_i$ of size $(\rho \times (\lambda\omega+\rho))$. These are built by making use of $V_{\ell}, \ell \in [2\lambda]$ and $(\lambda^2 + \lambda)$ diagonal matrices $\{ M_{ij} \ \mid \ i=1,2,\ldots, \lambda$,  $j=1,3,\ldots, 2\lambda-1$ or $j=2i\}$. If $j$ is odd, then $M_{ij}$ is of size $(\omega\times \omega)$ and otherwise, of size $(\rho\times \rho)$. We define $W_i, \ i \in [\lambda]$ as:
\bea 
W_i &\hspace{-2mm}\triangleq &\hspace{-2mm} \ [ W_{i,2i-1} \ W_{i,2i} \      W_{i,2i+1} \ \cdots \nonumber\\&&\  W_{i,2\lambda-1} \ W_{i,1} \ \cdots \   W_{i,2i-5} \   W_{i,2i-3}]~.~~~~
\eea 
where $W_{ij}=V_iM_{ij}$. The diagonal matrices $\{ M_{ij}\}$ are constructed in such a manner that each $W_i, \ i \in [\lambda]$ forms parity check matrix of a $[\lambda\omega+\rho, \lambda\omega, \rho+1]$ basic RS code. We also define these codes as:
\begin{equation}
{\cal D}_{[\lambda,\omega,\rho]}^{(i)}  =  \{ {\bf c} \in \mathbb{F}_q^{\lambda\omega+\rho} \mid W_i {\bf c}^T = {\bf 0} \}, \ i \in [\lambda] .
\end{equation} 
From the discussion in Sec.~\ref{sec:grs}, it may be observed that te diagonal entries of $M_{i1}, M_{i3},\ldots, M_{i,2\lambda-1}, M_{i,2i}$ are determined by $\Lambda_1 \cup \Lambda_3 \cup \cdots \cup \Lambda_{2\lambda-1} \cup \Lambda_{2i}$ for every $i$. It will be evident in the next step that these $\lambda$ basic RS codes form the local codes in the code ${\cal C}_{\text{BC}}[\mu,\lambda,\omega,\rho]$, that will be defined in \eqref{eq:cbc}. 

\item Next, we construct a $(\mu\rho \times \mu(\rho+\omega))$ parity check matrix $H_{\text{BC}}$. We first view $H_{\text{BC}}$ in a block-matrix form as in \eqref{eq:hbcblk}
\bea \label{eq:hbcblk}
H_{\text{BC}} & = & \left[ \small \begin{array}{cccc}
     H_{11} & H_{12} & \cdots & H_{1,2\mu}  \\
     H_{21} & H_{22} & \cdots & H_{2,2\mu}  \\
     \vdots & \vdots & \ddots & \vdots \\
     H_{\mu,1} & H_{\mu,2} & \cdots & H_{\mu,2\mu}  \\
\end{array} \right]_{\mu\rho \times \mu(\rho+\omega)}
\eea 
where $H_{ij}$ is of size $(\rho \times \omega)$ if $j$ is odd and of size $(\rho \times \rho)$ if $j$ is even. Then, we proceed to define each submatrix $H_{ij}, \ i \in [\mu], \ j \in [2\mu]$. Out of a total of $2\mu^2$ submatrices in \eqref{eq:hbcblk}, only $\mu(\lambda+1)$ submatrices are non-zero. For each $i=1,2,\ldots, \mu$, $H_{ij}$ is a non-zero matrix for $(\lambda+1)$ values of $j$ and these $(\lambda+1)$ non-zero submatrices are defined jointly in \eqref{eq:hbcnzero1} and \eqref{eq:hbcnzero2}. Let $i= s\lambda+t$ for $s \in \{0,1,\ldots, \nu-1\}$, $t \in \{1,2,\ldots, \lambda\}$. If $s < \nu-1$ or $(s = \nu-1, t=1)$, then 
\begin{equation}
\label{eq:hbcnzero1}
[H_{i,2i-1} \ H_{i,2i} \ H_{i,2i+1} \ H_{i,2i+3} \ \cdots \ H_{i,2(i+\lambda)-3}] \triangleq W_t .     
\end{equation}  
Or else, if $s = \nu-1$ and $t > 1$, then
\bea \label{eq:hbcnzero2}
&&\hspace{-10mm}[H_{i,2i-1} \ H_{i,2i} \  H_{i,2i+1} \ H_{i,2i+3} \ \cdots \nonumber\\
&&\ H_{i,2\mu-1} \ H_{i1} \ H_{i3} \ \cdots \ H_{i,2t-3}] \ := \  W_t.~~~~~ 
\eea
The remaining $2\mu^2-\mu(\lambda+1)$ submatrices are matrices with all zeros. This completes the definition of $H_{\text{BC}}$.

\item Finally, the code is defined as 
\bea \label{eq:cbc}
{\cal C}_{\text{BC}}[\mu,\lambda,\omega,\rho] & = & \{{\bf c}\in \mathbb{F}_q^n \mid H_{\text{BC}} {\bf c}^T = {\bf 0} \}.~~~~~
\eea 
\een 
When the parameters are clear in the context, we shall write ${\cal C}_{\text{BC}}$ in place of ${\cal C}_{\text{BC}}[\mu,\lambda,\omega,\rho]$. In $H_{\text{BC}}$, the submatrix formed by columns associated to $\{ W_{ij}, \text{$j$ is even}\}$ is of full rank. Therefore, the dimension of ${\cal C}_{\text{BC}}[\mu,\lambda,\omega,\rho]$ is equal to $k=\mu\omega$ and thus ${\cal C}_{\text{BC}}[\mu,\lambda,\omega,\rho]$ is a $[n=\mu(\rho+\omega), k=\mu\omega]$ code instantiating $T_{[\mu,\lambda,\omega]}(\rho)$. There are $\mu$ local codes ${\cal C}_{\text{BC},j}$ in ${\cal C}_{\text{BC}}$ defined by
\bean
{\cal C}_{\text{BC},i} & = & \{ {\bf c}|_{A_i} \ \mid \ {\bf c} \in {\cal C}_{\text{BC}} \} 
\eean 
where $A_i$ is as defined in Definition~\ref{def:bctop}. Observe that each of the $\mu$ local codes is one of $\lambda$ BRS codes ${\cal D}_{[\lambda,\omega,\rho]}^{(i)}, i \in [\lambda]$. In the next theorem, we characterize the minimum distance of the code \revp{under certain conditions}.

\bthm \label{thm:dmingt2} \revp{Let ${\cal C}_{\text{BC}}[\mu=\nu\lambda,\lambda,\omega,\rho]$ be a $[n=\mu(\rho+\omega), k =\mu\omega, d]$ linear code over $\mathbb{F}_q$ as defined in \eqref{eq:cbc}. If $\nu = 2^a$ for some integer $a \geq 0$ and $\mathbb{F}_q$ is of characteristic $2$, then 
\bea \label{eq:dmingt2}
d & = & \lambda\rho +1.
\eea }
\ethm
\bpf \revp{A sketch of the proof is given in Appendix~\ref{app:b}.} The full proof is available in a longer version of the paper \cite{SasVD24}.
\epf 

\section{An Application in Decentralized Systems \label{sec:bc}}


In this section, we describe the DA problem of blockchain networks, how linear codes play an important role in protocols addressing the DA problem, and finally how BC codes can serve as better alternatives to currently used 2D RS codes in certain operating regimes.

\subsection{The Data Availability Problem}
In a blockchain network, new blocks favored by consensus among a multitude of nodes are appended to a chain starting from the first block known as the genesis block. A block producer creates a block containing transactions picked up from a public pool and proposes it as the next block. \revp{Apart from transaction data, each block also has a header containing a hash of the previous block, which is typically of small size compared to the entire block size. Based on a consensus algorithm, the block is validated for the correctness of transactions and finalized in the chain. The consensus protocol ensures that only valid blocks are finalized and that a single chain survives among many forks. The initial architecture of the blockchain turned out to be non-scalable because every node would need to keep a local copy of the entire blockchain which can grow very large.} In response to this issue, a new type of node referred to as {\em light node} (in contrast to {\em full node} that was resource-rich) is introduced to achieve scalability. Light nodes only store the chain of block headers saving on the large memory requirement for transaction data. 

Light nodes verify transaction inclusion with the help of a cryptographic primitive called {\em vector commitment} (VC). Suppose that ${\bf t}=(\textsl{t}_1, \textsl{t}_2, \textsl{t}_3, \textsl{t}_4)$ represents the data within a block in a full node. Then $\textup{com}({\bf t})$, which is a {\em commitment} or {\em digest} of ${\bf t}$, is included in the block header. The number of bits in $\textup{com}({\bf t})$ is much less than that of ${\bf t}$. When a light node receives $\textsl{t}_1$ from the full node, 
it also obtains a small-sized proof $\pi(\textsl{t}_1)$ along with $\textsl{t}_1$. Given the proof and the digest $\textup{com}({\bf t})$ that it already has, the light node can verify if $\textsl{t}_1$ is a legitimate entry of ${\bf t}$. Although both $\textup{com}(\cdot)$ and $\pi(\cdot)$ are many-to-one functions, it is computationally hard to determine two different pre-images of a fixed value for both the functions. This forms the basis of security of a VC scheme. 

Although a light node can verify transaction inclusion in the above manner, it has to rely on an honest full node to validate the correctness of the transaction. If there is fraudulent transaction data, an honest full node will broadcast it with its proof. If a correct fraud-proof is received within a fixed amount of time, then the light node becomes aware of it and can reject the block header. The strategy works if light nodes are connected to at least one honest full node. However, if a malicious block producer publishes the block header but withholds a part of the transaction data, then the honest full node can not generate the fraud-proof. This is known as the {\em data availability} (DA) problem. A full node can not reject a block with incomplete data because, without the fraud-proof, it is impossible to distinguish whether the block producer is malicious or the challenger who detects the incompleteness of the block is malicious. 
A robust solution to the DA problem is that the light nodes ensure data availability in full nodes by themselves before accepting a block header (see \cite{But2018} for details.). In \cite{BasSBK2021}, Al Bassam et al. study the DA problem in depth and propose protocols based on random sampling and erasure coding as its solution. \revp{While the review in this and the next two subsections follows \cite{BasSBK2021}, a more abstract account from a cryptographic perspective can be found in \cite{HalSW23}}. 

\subsection{Data Availability Sampling with Erasure Code and Performance Criteria}

It defeats the purpose if a light node queries for the entire transaction data. Instead, it randomly samples a sufficiently small part of transaction data and this process is referred to as {\em data availability sampling} (DAS). The transaction data is split into $k$ chunks, encoded using an $[n,k,d]$ erasure code and the coded chunks are stored in full nodes. Because $(d-1)$ erasures can be recovered, a malicious node is forced to hide at least $d$ coded chunks to make the data unrecoverable. This enhances the probability that a light node queries for a hidden chunk. The block header will be accepted by the light node only if the queried chunks are returned with correct proofs. In the following, we describe the performance metrics used to evaluate the light node protocol as presented in \cite{BasSBK2021}.
Suppose that every block producer splits the entire transaction data into $k$ chunks, encodes it using an $[n,k,d]$ code, and includes these $n$ coded chunks into the block. Digests and proofs are also generated and included as part of coded data. Since each chunk of data is mapped to a symbol in a finite field, we use both chunk and symbol interchangeably. If there are $(n-d+1)$ chunks available, then the entire data is recoverable, and therefore the light nodes can accept the block header if there's no report of fraud. \revp{In addition, the light nodes must ensure that all missing chunks are recovered and every full node receives them. This guarantees the liveness of the system, i.e., every full node converges to the same correct state. On the other hand, if a block producer hides more than $(d-1)$ chunks, then the light nodes must reject the block header. This guarantees safety of the system, i.e.,  the header of a possibly fraudulent block that would later be detected by full nodes, does not get accepted by a honest light node.} 

\begin{figure*}
    \begin{center}
        \includegraphics[width=5.2in]{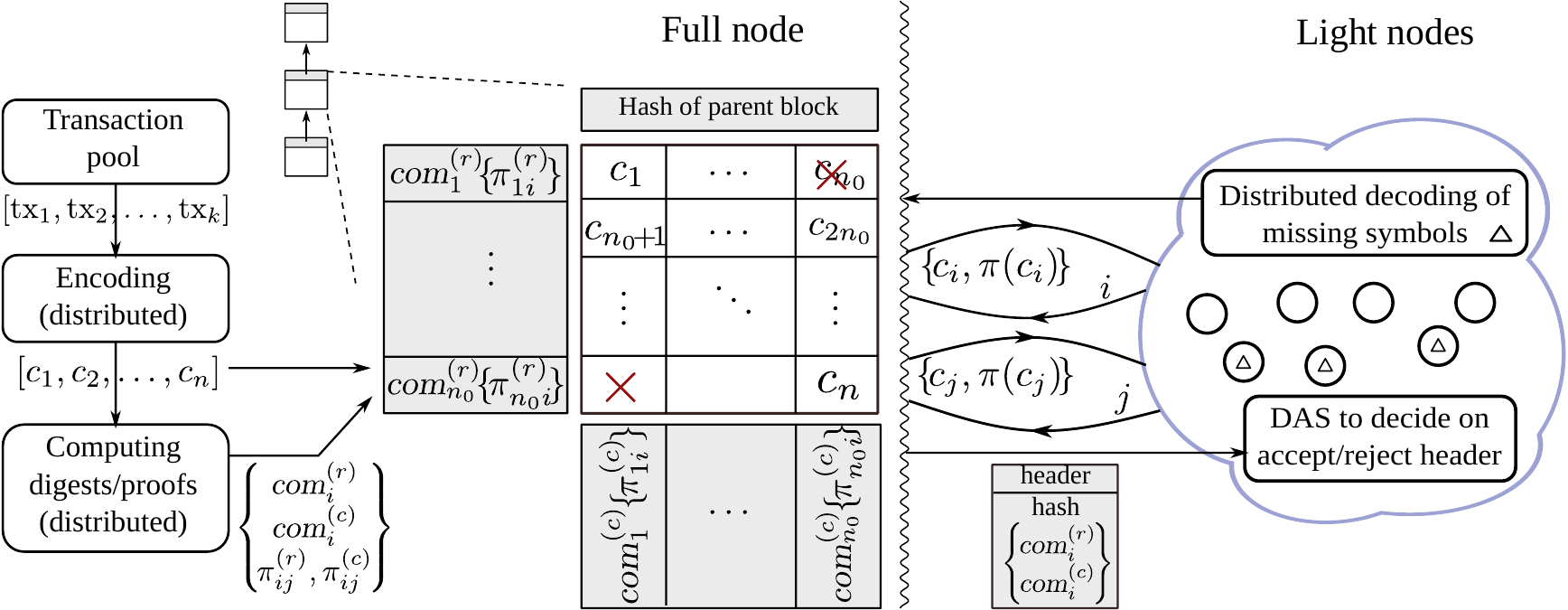}
        \caption{Illustration of the light node protocol (see Sec.~\ref{sec:prot}) based on 2D RS code, KZG commitments, and data availability sampling. Full nodes store transaction data ${\bf tx}=[\text{tx}_i,i=1,\ldots,k]$ as  $[c_i,i=1,\ldots,n]$ obtained by encoding ${\bf tx}$ by a $[n,k]$ 2D RS code. Block header includes KZG digests of every row/column codeword $\{\textsl{com}^{(r)}_i, \textsl{com}^{(c)}_i\}$, and KZG proofs $\{\pi^{(r)}_{ij}, \pi^{(c)}_{ij} \}$ of every symbol. Every light node queries fixed number of random symbols to decide whether to accept/reject a block header. Missing symbols can be decoded by a subset of light nodes (marked by $\triangle$) in a distributed manner.\label{fig:das}}
    \end{center}
    \end{figure*}
Assume that exactly $d$ chunks are hidden. Suppose a light node makes a DAS query of $0 < s < n-d+1$  distinct random chunks. Let $X$ denote the number of chunks that are unavailable among these $s$ chunks. The probability that the block header is rejected by the light node is the same as that of querying at least one unavailable chunk, i.e.
\bea \label{eq:p1s}
p_1(s) \ := \ \Pr (X \geq 1) & = & 1 - \prod_{i=0}^{s-1}\left( 1 - \frac{d}{n-i} \right) .~~~
\eea 
Suppose there are $c$ light nodes operating independently in the system, and let $Y$ denote the number of light nodes sampling at least one unavailable chunk. Then the probability that more than $c_0$ light nodes query for a  missing chunk is given by
\begin{equation} \label{eq:pcs}
p_c(c_0,s) :=  \Pr (Y > c_0)  =  1 - \sum_{j=0}^{c_0} {c \choose j} [p_1(s)]^j [1-p_1(s)]^{c-j} .
\end{equation} 
Let us define $\hat{c}(c, s, \gamma)$ as 
\bea  \label{eq:chat}
\hat{c}(c, s,\gamma) & = & \max_{1 \leq c_0 \leq c} \{ c_0 \mid p_c(c_0,s) \geq \gamma \} .
\eea 
for a threshold probability $\gamma$. When $\gamma = 0.99$, $\hat{c}(c, s, 0.99)$ is the maximum $c_0$ such that the probability $p_c(c_1,s)$ dips below $0.99$ for any $c_1 > c_0$. If $p_c(c_0,s) < \gamma$ for every $1 \leq c_0 \leq c$, then $(s,\gamma)$ is not achievable in this system and we have to either increase $s$ or decrease $\gamma$ to define $\hat{c}(c, s,\gamma)$ meaningfully.

Next, assume that there are at least $(n-d+1)$ chunks available. Then it is possible to do distributed decoding of every unavailable chunk if $(n-d+1)$ chunks are collectively sampled during DAS. The reconstructed samples are \revp{broadcasted} in the network of full nodes. Within some amount of time, all the honest full nodes will have the entire data available. In turn, every query of the light node will be returned. Let $Z$ denote the total number of distinct chunks out of $n$ chunks that are sampled. Then we can compute
\bea \label{eq:qcs}
&&\hspace{-6mm} q_c(s)  :=  \Pr( Z \geq n-d+1) =  \\ 
&& \hspace{-5mm}
1 - \sum_{i=1}^{n-d+1-s}\!\! (-1)^i {d+i-2 \choose d-1} {n \choose d+i-1} \left[ \frac{{n-d+1-i \choose s}}{{n \choose s}} \right]^c \!\!\!.\nonumber
\eea 
by invoking results from the theory of coupon collector's problem \cite{Stadje_1990}. We would like to have $q_c(s) \geq \eta$ where $\eta$ is approximately close to $1$ (say, $\eta=0.99$). Let us define
\bean
\label{eq:ctilde}
\tilde{c}(c,s,\eta) & = & \min_{1 \leq c_0 \leq c} \{ c_0 \mid q_{c_0}(s) \geq \eta \} ,
\eean 
for a given probability $\eta$. If $q_{c_0}(s) < \eta$ for every $1 \leq c_0 \leq c$, then $(s,\eta)$ is not achievable in this system, and we need to either decrease $\eta$ or increase $s$ to define $\tilde{c}(c, s, \eta)$ meaningfully. 

For a given value of $s$, we would like to have $\hat{c}(c, s,0.99)$ to be a large fraction of $c$ and $\tilde{c}(c,s,0.99)$ to be a small fraction of $c$. Turning it around, we can set performance targets on $\hat{c}(c, s,0.99)$ and $\tilde{c}(c,s,0.99)$, and ask for the minimum $s$ that achieves the targets. Given a performance target $(\hat{c}_T, \tilde{c}_T)$, we define
\bea \label{eq:smin}
s_{\text{min}} & = & \min \{ s \mid \hat{c}(c, s, \gamma) \geq \hat{c}_T , \ \tilde{c}(c,s,\eta) \leq \tilde{c}_T \}.~~~
\eea 
\revp{In order to reduce $s_{\text{min}}$, we need a code with a higher $(d/n)$, which will increase
the probability $p_1(s)$ in \eqref{eq:p1s}. At the same time, a larger $(k/n)$ reduces the storage overhead in full nodes. In addition, it must also satisfy Properties $1,2$ and $3$ for reasons described in Sec.~\ref{sec:intro}. }

\subsection{\revp{Light Node Protocol Using 2D Reed-Solomon Code}\label{sec:prot}}

\revp{An $[n=n_0^2,k=k_0^2,d=d_0^2]$ 2D RS code instantiating the square product topology with $[n_0,k_0,n_0-k_0+1]$ row/column codes satisfies Properties $1,2$ and $3$. The complete protocol with 2D RS code is developed in \cite{BasSBK2021}. It is illustrated in Fig.~\ref{fig:das} and works as follows:}
\ben 
\item \revp{The full node will have encoded transaction data, commitments $\{\textsl{com}^{(r)}_i,\textsl{com}^{(c)}_i \mid i \in [n_0]\}$ associated to every row/column local codewords, and associated proofs for every code symbol $\{\pi^{(r)}_{ij}, \pi^{(c)}_{ij} \mid i \in [n_0], j\in [n_0]\}$ as shown in Fig.~\ref{fig:das}. The proof $\pi(c)$ associated to a particular symbol $c$ consists of proofs with respect to both row and column digests.}
\item Upon receiving a block header, every light node samples $s \geq s_{\text{min}}$ chunks independently of any other, accepts the header if the samples are returned with correct proof within a fixed amount of time, and rejects it otherwise.  
\item If $c > \tilde{c}_T$, they will successfully query $n-d+1$ chunks collectively with a high probability of at least $\eta$. Distributed decoding of missing chunks can be carried out by a subset of light nodes (as marked in Fig.~\ref{fig:das}), and the reconstructed chunks are gossiped among all full nodes. All the chunks will be available within a fixed amount of time and hence queries of light nodes will be successfully returned with high probability in that time. Eventually, this leads to acceptance of block header.
\item On the other hand, if more than $d$ chunks are hidden in an adversarial manner, the distributed decoding algorithm fails to recover unavailable chunks. Hence, queries of more than $\hat{c}_T$ light nodes will fail with a high probability of at least $\gamma$, causing rejection of block header. 
\een 

\revp{The 2D RS code performs well in the low-rate regime. For example in the initial proposal of $k_0=(n_0/2)$ i.e., with an expensive $4$x storage overhead at full node, $(d/n) \approx 25\%$ which is quite high for practical purposes\cite{BasSBK2021}. The code does not require resource-rich `super-nodes' dedicated to decoding and encoding\cite{Fei22_YT}. Larger values of $k_0$ are considered in \cite{SanRBCB22}, but cause a sharp dip in $(d/n)$. We propose replacing the 2D RS code with a BC code, offering benefits in the high-rate regime, as detailed later in Sec.~\ref{sec:bcvs2drs}}. 

\begin{figure*}
     \centering
     \subfloat[][$p_1(s)$ versus $s$.]{\includegraphics[width=2.2in]{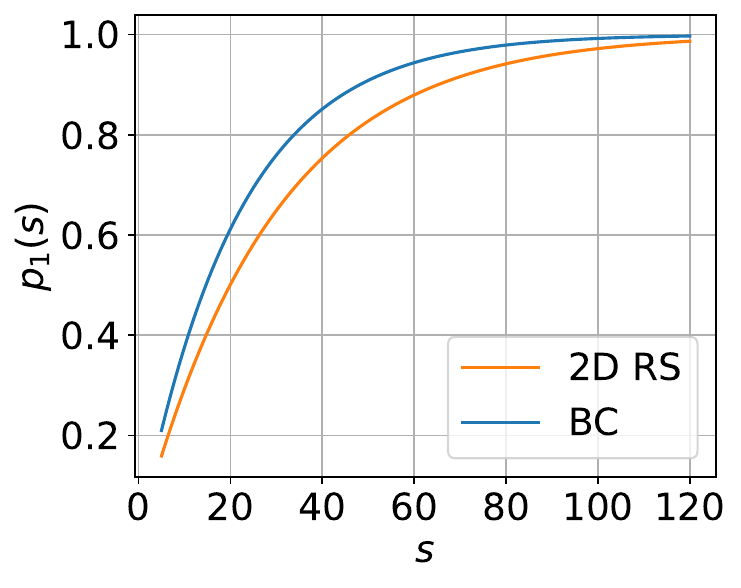}\label{fig:daplot1}}
     \subfloat[][$\hat{c}(c,s,\gamma)$ versus $c$.]{\includegraphics[width=2.3in]{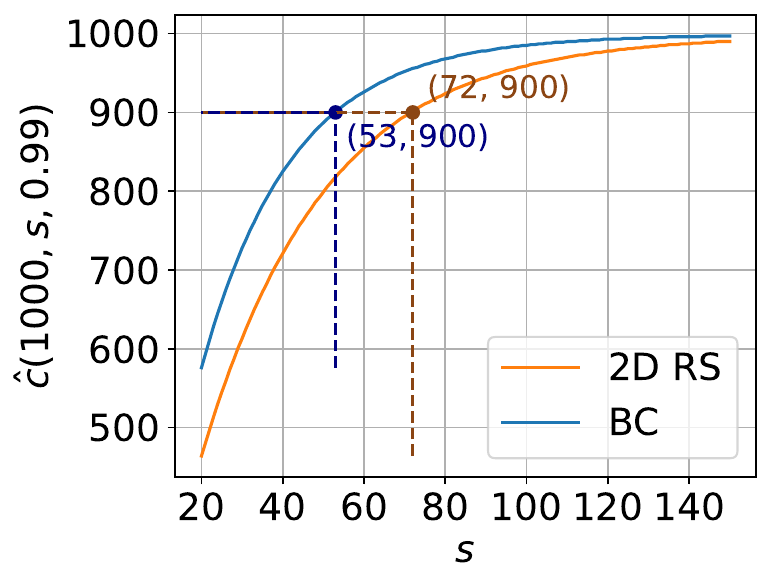}\label{fig:daplot2}}
     \subfloat[][$q_c(s)$ versus $c$.]{\includegraphics[width=2.2in]{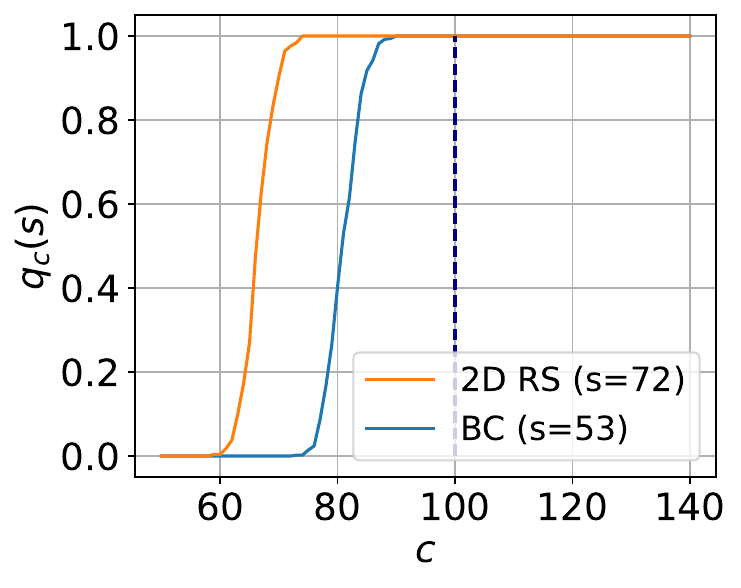}\label{fig:daplot3}}     
     \caption{Performance of data availability sampling with $[1444,1024,49]$ 2D RS and $[1416,1024,65]$ BC codes. We set $c=1000, \gamma=\eta=0.99$.}
     \label{fig:daplot}
\end{figure*}

\subsection{Two Variants of Commitement Schemes \label{sec:variants}}

\revp{There are two variants of the light node protocol based on which vector commitment scheme is used. The first denoted as ${\cal P}_{\text{Mer}}$ uses the VC scheme based on Merkle trees, though it may be realized with any arbitrary VC scheme. The protocol is developed and integrated as part of a working system \cite{Al2019}.}

A security risk in the protocol as described in Sec.~\ref{sec:prot} is that the coded symbols may be incorrectly generated in the first place. This is called the {\em incorrect coding (IC) attack}. 
An honest full node can detect an incorrectly coded chunk by detecting that a p-c constraint associated to at least one $[n_0,k_0,d_0]$ row- or column-code $C_{\text{cfp}}$ fails. Then it broadcasts a fraud-proof, referred to as {\em codec fraud-proof (CFP)}, among light nodes. The CFP consists of (a) an index of $C_{\text{cfp}}$, (b) $k_0$ chunks belonging to $C_{\text{cfp}}$,  and (c) proofs for these chunks with respect to the digest corresponding to $C_{\text{cfp}}$. Upon receiving a block header, every light node, in addition to making queries for $s$ random samples, will also wait for a fixed amount of time $T_{\text{cfp}}$ for the arrival of any CFP. A CFP can be validated using digest associated to $C_{\text{cfp}}$. The block header will be accepted only if no correct CFPs are received, in addition to earlier conditions. The Property $1$ of the code helps to reduce the size and computational complexity of CFPs. Later, a framework called coded Merkle tree (CMT) \cite{YuSLAKV20} based on LDPC codes, and avoiding the use of 2D RS code, was introduced to reduce CFP size to an optimal minimum. Optimizations on LDPC code used for CMT were then carried out in \cite{MitTD22}. A secure distributed decoding protocol for LDPC codes was proposed in \cite{CaoKR2020}. In \cite{SheXKV2021}, a coded interleaving tree (CIT) was proposed to avoid the need of anonymous communication channels required in CMT. 

\revp{The second variant denoted as ${\cal P}_{\text{KZG}}$ uses Kate-Zaverucha-Goldberg (KZG) polynomial commitment scheme. It evolved later mostly as a result of industry research and is being pushed into practice by Ethereum~\cite{ER2023,WagZ24}.} The KZG VC scheme proposed in \cite{KateZG10} is specifically intended for vectors whose entries are evaluations of a fixed polynomial. Given the digest of a vector ${\bf c}=(c_i, \ i=1,\ldots, n_0)$ and proof for $c_i$, the KZG scheme can verify if $c_i$ is part of ${\bf c}$ and if $c_i$ is an evaluation of a fixed polynomial associated to ${\bf c}$. The 2D RS code made up of polynomial evaluation local codes is commitment-compatible with the KZG scheme. Thus, ${\cal P}_{\text{KZG}}$ does not require CFPs and handles the IC attack most efficiently. \revp{The 2D RS code is a unique code known in literature satisfying all three properties, and at the same time compatible with KZG scheme. The BC code that we propose is therefore the second of its kind.} 

\subsection{BC Code as an Alternative to 2D RS Code \label{sec:bcvs2drs}}
 
\revp{In both ${\cal P}_{\text{Mer}}$ and ${\cal P}_{\text{KZG}}$, replacing the 2D RS code with a BC code enables expansion to a broader range of parameter sets. In particular, the following example shows that a BC code can outperform the 2D RS code in $(d/n)$ and other metrics in the high-rate regime. }

\revp{We pick two codes, one from the 2D RS family and another from the BC family, both having approximately the same storage overhead $(n/k) \approx 1.4$x for a fair comparison. Both have $k=1024$, so the size of a data chunk remains fixed for a fixed block size. The 2D RS code with parameters $[1444, 1024, 49]$ is denoted by ${\cal C}_{\text{2DRS}}[32]$. The $[n=\mu(\rho+\omega)=1416,k=\mu\omega=1032, d=\lambda\rho+1=65]$ BC code ${\cal C}_{\text{BC}}[12,2,86,32]$ is shortened by $8$ symbols to obtain a $[1416, 1024, 65]$ code, denoted as ${\cal C}_{\text{BC}}^{\mtiny{(8)}}[12,2,86,32]$. Parameters of the codes and their local codes are listed in Table~\ref{tab:comp1}. The BC code has a higher $(d/n)$ for the same rate. Although both codes are realizable over $\mathbb{F}_{256}$, larger prime fields are typically used for implementing the KZG scheme.}
\setlength{\tabcolsep}{4pt}
\begin{table}[h]
	\centering 
		\revp{\begin{tabular}{|c|c|c|c|c|}
    \hline 
    Code & $[n,k,d]$ & $[n_0,k_0,d_0]$ & $(n/k)$ & $(d/n)$ 
  \\
 		\hline \hline 
                 ${\cal C}_{\text{2DRS}}[32]$ & $[1444,1024,49]$ & $[38,32,7]$ & $1.4$x & $3.39\%$ \\
                \hline 
                 ${\cal C}_{\text{BC}}^{\mtiny{(8)}}[12,2,86,32]$ & $[1416,1024,65]$& $[204,172,33]$ & $1.4$x & $4.59\%$ \\
                 \hline 
    \end{tabular}} 
    \caption{Parameters of example 2DRS and BC codes. \label{tab:comp1}} \vspace{-4mm}
\end{table} 
\setlength{\tabcolsep}{3pt} 
\begin{table}[h]
	\centering 
		\revp{\begin{tabular}{|c|c|c|c|c|c|c|}
    \hline 
     &  & \# digests & \multicolumn{2}{c|}{${\cal P}_{\text{Mer}}$} & ${\cal P}_{\text{KZG}}$ & \# chunks  \\
    \cline{4-5} 
    Code & $s_{\text{min}}$ 
 & to & Header & CFP & Header & per node 
 \\
     & & compute & size & size & size & (decoding) \\
			\hline \hline 
                 ${\cal C}_{\text{2DRS}}[32]$ & $72$ & $77$ & $77$ & $32$ & $65$ & $32$ \\
                \hline 
                 ${\cal C}_{\text{BC}}^{\mtiny{(8)}}[12,2,86,32]$ & $53$ & $13$ & $13$ & $172$ &  $13$ & $172$ \\
                 \hline 
    \end{tabular}} 
    \vspace{0.1in}
	\caption{Comparison of metrics related to the light node protocol. The most relevant metric $s_{\text{min}}$ is lower for the BC code. We set $c=1000$, $\gamma = \eta = 0.99$, and a target performance of $(\hat{c}_T=900, \tilde{c}_T=100)$.  \label{tab:comp2}} \vspace{-4mm}
\end{table} 

\revp{In order to evaluate the performance of two codes in terms of metrics related to the light node protocol, let us consider a set of $c=1000$ light nodes and set $\gamma=\eta=0.99$. Suppose we aim to achieve a target of $\hat{c}(1000,s,0.99)=900$ (i.e., $90\%$ of all light nodes) and $\tilde{c}(1000,s,0.99)=100$ (i.e., $10\%$ of all light nodes). In Fig.~\ref{fig:daplot1}, $p_1(s)$ is plotted for both codes, and the BC code clearly performs better because of its larger $(d/n)$. In Fig.~\ref{fig:daplot2}, the minimum $s$ to achieve $\hat{c}(1000,s,0.99)=900$ is determined under both codes. We obtain $s_{\text{min}}=72$ for ${\cal C}_{\text{2DRS}}[32]$ and $s_{\text{min}}=53$ for ${\cal C}_{\text{BC}}^{\mtiny{(8)}}[12,2,86,32]$, yielding a significant reduction in network traffic for the BC code. In Fig.~\ref{fig:daplot3}, it is validated that the above $s_{\text{min}}$ is sufficient to achieve the second target $\tilde{c}(1000,s,0.99)=100$. }

\revp{There are $L=76$ local codes in ${\cal C}_{\text{2DRS}}[32]$ each of dimension $k_0=32$, whereas $L=12$ in ${\cal C}_{\text{BC}}^{\mtiny{(8)}}[12,2,86,32]$ with $k_0=172$. Implementing the commitment scheme requires pre-computation of digests one each for every local codeword and one for the entire codeword. The number of digests to be computed therefore drops to $13$ in ${\cal C}_{\text{BC}}^{\mtiny{(8)}}[12,2,86,32]$, compared to $77$ in ${\cal C}_{\text{2DRS}}[32]$. The size of the block header is proportional to the number of digests. However in ${\cal P}_{\text{2DRS,KZG}}$, the homomorphic property of KZG scheme\cite{But2020} can be exploited to compute digests of the remaining local codes on the fly, if those of $64$ local codes are precomputed and stored. This allows to store only $65$ digests in the header for ${\cal P}_{\text{2DRS,KZG}}$ instead of $77$ required for ${\cal P}_{\text{2DRS,M}}$. But replacement with BC code in ${\cal P}_{\text{2DRS,M}}$ and ${\cal P}_{\text{2DRS,KZG}}$ reduces the number of digests, and hence header size significantly to $13$. We remark that the time complexity to compute the KZG digest of a local codeword is linear in $k_0$. Thus, the net time complexity of computing digests grows as $Lk_0$, which is reduced by a constant factor for the BC code. }

\revp{The size of CFP in ${\cal P}_{\text{2DRS,M}}$, the number of symbols to be accessed by a node taking part in distributed decoding are proportional to $k_0$. Every node participating in distributed decoding decodes a $[38,32]$ GRS code for ${\cal C}_{\text{2DRS}}[32]$, whereas it is either a $[204,172]$ or a $[236,172]$ GRS code in ${\cal C}_{\text{BC}}^{\mtiny{(8)}}[12,2,86,32]$, requiring larger per-node time-complexity. Thus the ${\cal C}_{\text{BC}}^{\mtiny{(8)}}[12,2,86,32]$ suffers compared to ${\cal C}_{\text{2DRS}}[32]$ in these three metrics while gaining in all others. All the protocol-related metrics are tabulated in Table~\ref{tab:comp2}.}

\section{Conclusion\label{sec:con}}
In this paper, we proposed block circulant codes ${\cal C}_{\text{BC}}[\mu,\lambda,\omega,\rho]$ that are fully defined by local p-c constraints just like the product codes, but outperform square product codes in relative minimum distance in the high rate regime ($\geq 0.5$). The code has an efficient, parallelizable erasure-decoding algorithm when the overlap factor $\lambda =2$. 
The block circulant topology of the code makes it suitable for light node protocols in blockchain networks. Analysis suggests that BC codes with $\lambda=2$ work better than currently proposed 2D RS codes of the same storage overhead in tackling the DA problem in blockchain networks. A prototype implementation to validate the same and further fine-tuning by exploiting larger values of $\lambda$ may be taken up as future work. 



\bibliographystyle{IEEEtran}
\bibliography{bc.bib}
\appendices
\section{\revp{Proof of Theorem~\ref{thm:dmin2}}\label{app:a}}

To prove the assertion $(1)$, we will show that the algorithm terminates and that $r_j, j=0,1,\ldots, n-1$ is either decoded correctly or unchanged during the course of the algorithm. 

Recall that ${\cal C}_{\text{BC}}[2]$ consists of $\mu$ local codes ${\cal C}_{\text{BC},i}[2], i=1,2,\ldots, \mu$ and each ${\cal C}_{\text{BC},i}[2]$ is a $[2\omega+\rho, 2\omega, \rho+1]$ BRS code that can correct any pattern of up to $\rho$ erasures. Moreover, ${\cal C}_{\text{BC},i}[2]={\cal D}^{(1)}_{[2,\omega,\rho]}$ when $i$ is odd, and ${\cal C}_{\text{BC},i}[2]={\cal D}^{(2)}_{[2,\omega,\rho]}$ when even. We have correct erasure-correcting decoders $\textsc{DEC}_{{\cal D}^{(i)}[2]}$ for ${\cal D}^{(i)}_{[2,\omega,\rho]}, i=1,2$. In \textsl{Line} $4$, the algorithm computes a collection $J_1$ of all local codes whose codewords contain at least $1$ and at most $\rho$ erased symbols. Every erasure within these local codes are corrected (see \textsl{Lines} $6$-$8$), and then the set $J_1$ is recomputed again (see \textsl{Line} $10$). It may so happen that a few other local codes qualify to be within $J_1$ at this point. Therefore, the procedure is repeated in a loop starting at \textsl{Line} $5$, until $J_1$ becomes empty. Since at least one erasure is decoded in every iteration of the loop, and that $J_1$ is updated after every iteration, it is clear that the loop at \textsl{Line} $5$ must end in a finite number of iterations. At (say) $t$-th iteration, symbols recovered by correct decoding of different local codes have been merged into a single vector ${\bf \hat{r}}[t+1]$. Therefore, the vector ${\bf \hat{r}}$ set as ${\bf \hat{r}}[t_f]$ in \textsl{Line} $13$ where $t_f$ is the final iteration will include all symbols that have been recovered over all iterations. Thus, after the execution of \textsl{Line} $13$, every $\hat{r}_j, j=0,1,\ldots,n-1$ in ${\bf r}$ is either correctly decoded or unchanged.

Consider that \textsl{Line} $13$ is executed. Over \textsl{Lines} $14$-$25$, the vector ${\bf \hat{r}}$ remains unchanged. A set $J_2$ is computed, separately handling two exhaustive cases $\mu=2$ and $\mu \geq 4$. Let us define
\begin{equation} \label{eq:J}
    J = \cup_{(i,i')\in J_2} \{i,i'\} .
\end{equation} 
If there is an erasure in ${\bf \hat{r}}_{A_i}$ where $i \notin J$, then it is not corrected by the algorithm. In that case, the algorithm declares that ``{\bf r} has uncorrectable erasures.'' Every erasure in ${\bf \hat{r}}_{A_i}, i \in J$ are correctly recovered, as will be argued below. 

Consider that \textsl{Line} $25$ is executed. Since $|J_2| \leq \mu$, the loop at \textsl{Line} $26$ must have only a finite number of iterations. Consider one iteration of the loop associated to $(i_1,i_2=(i_1+1)_{\mu})\in J_2$. Let us consider the BRS local codewords ${\bf c}_1 := {\bf c}|_{A_{i_1}}$ and ${\bf c}_2 := {\bf c}|_{A_{i_2}}$. We also write ${\bf c}_i = [{\bf c}_{i1} \ {\bf c}_{i2} \ {\bf c}_{i3}]$ where ${\bf c}_{i1}, {\bf c}_{i3} \in \mathbb{F}_q^{\omega}$ and ${\bf c}_{i2} \in \mathbb{F}_q^{\rho}$ for $i=1,2$. We observe that ${\bf c}_{13}={\bf c}_{21}$. If $i_1$ is odd, then ${\bf c}_{11},{\bf c}_{23}$ are associated with set of code locators $\Lambda_1$, ${\bf c}_{13},{\bf c}_{21}$ with $\Lambda_3$, ${\bf c}_{12}$ with $\Lambda_2$ and ${\bf c}_{22}$ with $\Lambda_4$. If $i_1$ is even, then we need to interchange $\Lambda_1$ with $\Lambda_3$, and $\Lambda_2$ with $\Lambda_4$ in the previous statement. Since every argument of the proof with such an interchange still remains true, we assume that $i_1$ is odd without loss of generality. Let $m_1(X), m_2(X)$ denote the message polynomials of degree at most $2\omega-1$ associated to ${\bf c}_1$ and ${\bf c}_2$. 

We will first establish that the polynomial $\hat{s}(X)$ computed in \textsl{Line} $32$ is identically equal to $s(X) \triangleq m_1(X)-m_2(X)$. If $\mu=2$, then $(i_1,i_2)=(1,2)$. Furthermore, 
\bea \label{eq:mu2s1}
(m_1(\alpha), \alpha \in \Lambda_{1}) = {\bf c}_{11} \!\! & = & \!\! {\bf c}_{23} = (m_2(\alpha), \alpha \in \Lambda_{1}), \\
\label{eq:mu2s2} (m_1(\alpha), \alpha \in \Lambda_{3}) = {\bf c}_{13} \!\! & = & \!\! {\bf c}_{21} = (m_2(\alpha), \alpha \in \Lambda_{3}).
\eea 
Hence $m_1(X)$ and $m_2(X)$ must be the same because they evaluate the same on $2\omega$ distinct points. Thus $s(X)$ is the zero polynomial, which equals $\hat{s}(X)$ as computed in \textsl{Lines} $27$ and $33$. Next, consider the case $\mu \geq 4$. We have $((m_1-m_2)(\alpha), \alpha \in \Lambda_3) = (m_1(\alpha), \alpha \in \Lambda_3) - (m_2(\alpha), \alpha \in \Lambda_3) = {\bf c}_{13}-{\bf c}_{21}={\bf 0}$. We also have 
\bea
\nonumber ((m_1-m_2)(\alpha), \alpha \in \Lambda_1) & = & {\bf c}_{11}-{\bf c}_{23} \\
\label{eq:pf1} & = & {\bf \hat{r}}|_{B_{i_1,1}} - {\bf \hat{r}}|_{B_{i_2,2}}.
\eea  
The equation \eqref{eq:pf1} is true because there are no erasures in ${\bf \hat{r}}$ within $B_{i_1,1} \cup B_{i_2,2}$ by the definition of $J_2$. Therefore, ${\bf \hat{s}}$ computed in \textsl{Line} $30$ is a local codeword of ${\cal D}^{(1)}_{[2,\omega,\rho]}$, but with $\rho$ symbols erased. The decoding in \textsl{Line} $31$ will correct all the $\rho$ erasures, and hence $\hat{s}(X)$ in \textsl{Line} $33$ is equal to $s(X)$. 

Next, we proceed to show that vectors ${\bf \hat{u}}_1, {\bf \hat{u}}_2$ computed in \textsl{Lines} $37$-$38$ are respectively $(m_1(\alpha), \alpha \in \Lambda )$ and $(m_2(\alpha), \alpha \in \Lambda)$ where $\Lambda = \cup_{\ell=1}^4 \Lambda_{\ell}$. We have $m_1(X) = m_2(X)+s(X)$ and $m_2(X) = m_1(X)-s(X)$. Since $s(X)$ is known, we obtain $m_1(\alpha), \alpha \in \Lambda$ if it is unerased in ${\bf c}_1$ or if $m_2(\alpha)$ is unerased in ${\bf c}_2$. In similar lines, we obtain $m_2(\alpha)$ as well in two different ways. Therefore, the vector ${\bf \hat{u}}_i, i=1,2$ computed in \textsl{Lines} $35$-$36$ has evaluations of $m_i(\alpha)$ for every $\alpha \in \Lambda$, except when both $m_1(\alpha)$ and $m_2(\alpha)$ are erased. But, it is clear that
\begin{align}
\nonumber |\{ \alpha \in \Lambda \mid m_1(\alpha) \text{ and } m_2(\alpha) & \text{ are unavailable at \textsl{Line} $34$} \}| \\
& \leq    n_E({\bf \hat{r}}|_{A}) \ \leq \ 2\rho,    
\end{align}
where $A \triangleq A_{i_1} \cup A_{i_2}$. Therefore, $n_E({\bf \hat{u}}_i) \leq 2\rho$ for $i=1,2$. Since ${\bf \hat{u}}_i, i=1,2$ are codewords of the $[2(\omega+\rho),2\omega,2\rho+1]$ auxiliary code ${\cal D}^{(1\cup 2)}_{[2,\omega,\rho]}$, \textsl{Lines} $37$-$38$ will correctly recover from all erasures in ${\bf \hat{u}}_i, i=1,2$. 

In \textsl{Line} $39$, all erasures in ${\bf \hat{r}}|_A$ are updated with appropriate correct values of $m_i(\alpha),\alpha \in \Lambda$ where $i \in [2]$. Therefore every symbol lying on $A_i, i \in J$ are correctly recovered over the loop at \textsl{Line} $26$, and symbols that do not fall within $A_i, i \in J$ are unchanged. This completes the proof of assertion $(1)$.

To show assertion (2), consider an input {\bf r} such that $n_E({\bf r}) \leq 2\rho$. If $0 < n_E({\bf r}) \leq \rho$, then clearly $\{i \mid n_E({\bf r}|_{A_i}) > 0 \} \subseteq J_1$ (computed in \textsl{Line} $4$). This implies that all erasures are corrected by the time the algorithm reaches \textsl{Line} $13$. Suppose $\rho < n_E({\bf r}) \leq 2\rho$ and $n_E({\bf \hat{r}}) > 0$ where ${\bf \hat{r}}$ corresponds to the state immediately after executing \textsl{Line} $13$. Then, we claim that $\{i \mid n_E({\bf \hat{r}}|_{A_i}) > 0 \} = J$. If $\mu =2$, the claim is trivially true. So assume that $\mu \geq 4$ and the claim is false. Then there exists $i, i'$ such that $i' \notin \{(i +1)_{\mu},(i -1)_{\mu}\}$ and $n_E({\bf \hat{r}}|_{A_i}) > \rho$ and $n_E({\bf \hat{r}}|_{A_{i'}}) > \rho$. This implies that $n_E({\bf \hat{r}}) \geq 2\rho+2 > 2\rho+1$. Hence $n_E({\bf r}) \geq n_E({\bf \hat{r}}) > 2\rho$, a contradiction to the assumption. This proves the claim. Since symbols lying in $A_i, i \in J$ are all correctly decoded, ${\bf \hat{r}} = {\bf c}$ when the algorithm terminates. This completes the proof.

\section{\revp{Proof Sketch of Theorem~\ref{thm:dmingt2}}\label{app:b}}

The first row of a systematic generator matrix of the code (constructed in similar lines as how it is done for $\lambda=2$ in Sec~\ref{sec:enc2}) is of weight $\lambda\rho+1$. Therefore, $d \leq \lambda\rho+1$. The tedious part of the proof is to show that $d \geq \lambda\rho+1$. Let ${\bf c}$ be an arbitrary non-zero codeword in ${\cal C}_{\text{BC}}[\mu,\lambda,\omega,\rho]$. Our approach is to construct a polynomial $f(X; {\bf c})$ of degree at most $(\lambda\omega-1)$, based on ${\bf c}$ as the notation emphasizes. First, we consider the case that $f(X;{\bf c})$ is a non-zero polynomial. In that case, we identify $\lambda(\rho+\omega)$ evaluations of $f(X;{\bf c})$ as linear combinations of symbols in ${\bf c}$. Since $f(X;{\bf c})$ must evaluate to non-zero values at least $\lambda\rho+1$ values, it would imply that the codeword also must have at least that many non-zero symbols. Then, we consider the case that $f(X;{\bf c})$ is a zero polynomial. In that case, we use induction on $a$ where $\nu = 2^a$ to prove that $w_H({\bf c}) \geq \lambda\rho+1$. The induction works only when the field is of characteristic $2$. This would complete the proof. The following sequence of arguments provides a sketch of the proof that $w_H({\bf c}) \geq \lambda\rho+1$ under conditions specified in the theorem.
\ben
\item We construct $\lambda$  polynomials $f_i(X; {\bf c}) = \sum_{j=0}^{\nu-1} m_{i+j\lambda}(X; {\bf c}), \ i \in [\lambda]$ each of degree $\lambda\omega-1$, where $m_i(X; {\bf c})$ denote the message polynomial associated to local codeword ${\bf c}|_{A_i} \in {\cal C}_{\text{BC},i}, i \in [\mu]$. 
\item We identify that $f_i(X;{\bf c}), i \in [\lambda]$ are identically the same (hence, denote it as $f(X;{\bf c})$) because they all evaluate to the same values at $\lambda\omega$ distinct points in the set $\Lambda_{\text{odd}} := \cup_{\ell_0=1}^{\lambda} \Lambda_{2\ell_0-1}$. If we write ${\bf c} = [{\bf c}_1 \ {\bf c}_2 \ \cdots \ {\bf c}_{2\mu}]$ where ${\bf c}_j \in \mathbb{F}_q^{\omega}$ for odd $j$ and ${\bf c}_j \in \mathbb{F}_q^{\rho}$ for even $j$, then for $\ell=1,3,\ldots,2\lambda-1$,
\bea \label{eq:fevalodd}  
f_i(\Lambda_{\ell};{\bf c}) & = & {\bf c}_{\ell} + {\bf c}_{\ell+\lambda} + \cdots + {\bf c}_{\ell+(\nu-1)\lambda} .
\eea 
\item Next, the line of arguments splits into two exclusive and exhaustive cases, viz., that $f(X;{\bf c})$ is a non-zero polynomial as {\em Case 1} and otherwise as {\em Case 2}. In {\em Case 1}, we obtain evaluations of $f(X;{\bf c})$ in $\Lambda = \Lambda_{\text{odd}} \cup \Lambda_{\text{even}}$ 
where $\Lambda_{\text{even}} := \cup_{\ell_0=1}^{\lambda} \Lambda_{2\ell_0}$. For $\ell = 2,4,\ldots, 2\lambda$, we compute
\bea 
\nonumber f(\Lambda_{\ell};{\bf c}) & = & f_{(\ell/2)}(\Lambda_{\ell};{\bf c}) = \sum_{j=0}^{\nu-1} m_{(\ell/2)+j\lambda}(\Lambda_{\ell}; {\bf c})  \\
\label{eq:fevaleven} & = & {\bf c}_{\ell} + {\bf c}_{\ell+\lambda} + \cdots + {\bf c}_{\ell+(\nu-1)\lambda},
\eea 
and \eqref{eq:fevalodd} and \eqref{eq:fevaleven} collectively provide $\lambda(\omega+\rho)$ distinct evaluations of $f(X;{\bf c})$. Since $f(X;{\bf c})$ is of degree at most $(\lambda\omega-1)$, at least $\lambda\rho+1$ of the above evaluations of must be non-zero. 
\item We express \eqref{eq:fevalodd} and \eqref{eq:fevaleven} jointly as
\bean
f(\Lambda;{\bf c}) = \sum_{j=0}^{\nu-1} [{\bf c}_{1+2j\lambda}, {\bf c}_{2+2j\lambda},\cdots, {\bf c}_{2\lambda+2j\lambda} ] 
\eean 
if points in $\Lambda$ are suitably ordered. Then we have,
\bea
\label{eq:evalsumwt} \sum_{j=0}^{\nu-1} w_H([{\bf c}_{1+2j\lambda}, {\bf c}_{2+2j\lambda},\cdots, {\bf c}_{2\lambda+2j\lambda} ]) \geq \lambda\rho+1 ,
\eea 
leading to the conclusion that $w_H({\bf c}) \geq \lambda\rho + 1$.
\item When $\nu=1$, i.e., $\mu = \lambda$, $f(X;{\bf c})$ must be non-zero because $f(X;{\bf c}) = f_i(X;{\bf c})=m_i(X;{\bf c})$ for every $i \in [\lambda]$ and at least one of $m_i(X)$ must be non-zero when ${\bf c} \neq {\bf 0}$. Thus, the proof for $\nu=1$ is subsumed in {\em Case 1}.
\item In  {\em Case 2}, the proof that $w_H({\bf c}) \geq \lambda\rho+1$ is by induction on $a \in \mathbb{N}$ where $\mu=\nu\lambda = 2^{a}\lambda$. In the base case of $a=1$, $\mu = 2^a\lambda = 2\lambda$, let us consider a non-zero vector ${\bf z}_1  \in {\cal C}_{\text{BC}}[2\lambda,\lambda,\omega,\rho]$. Since $f(X;{\bf z}_1)=0$, we must have
\bea
\nonumber m_i(X;{\bf z}_1) + m_{i+\lambda}(X;{\bf z}_1)  & = &  0,\\
\label{eq:f0basemequal} \Rightarrow m_i(X;{\bf z}_1) & = &  m_{i+\lambda}(X;{\bf z}_1),
\eea 
for $i = 1,2,\ldots, \lambda$ because $\mathbb{F}_q$ is of characteristic $2$.
We write ${\bf z}_1 = [{\bf z}_{11} \ {\bf z}_{12} \cdots {\bf z}_{1,4\lambda}]$ where ${\bf z}_{1,j}$ is a $(1\times \omega)$ vector if $j$ is odd and a $(1 \times \rho)$ vector if $j$ is even. By \eqref{eq:f0basemequal}, we must have
\begin{equation} \label{eq:f0basezequal}
{\bf z}_{1,j} = {\bf z}_{1,j+2\lambda}, \ j = 1,2,\ldots, 2\lambda.
\end{equation} 
Let us define ${\bf z}^{\mtiny{(1)}}_1 = [{\bf z}_{11} \ {\bf z}_{12} \cdots {\bf z}_{1,2\lambda}]$ and ${\bf z}^{\mtiny{(2)}}_1 = [{\bf z}_{1,2\lambda+1} \ {\bf z}_{1,2\lambda+2} \ \cdots \ {\bf z}_{1,4\lambda}]$. It follows from \eqref{eq:f0basezequal} that any parity check constraint involving ${\bf z}_{1,j+2\lambda}$ can be rewritten by replacing ${\bf z}_{1,j+2\lambda}$ by ${\bf z}_{1,j}$ and vice versa. This leads to the equation
\bea \label{eq:z1split}
\left[ 
\begin{array}{cc}
	H_{\text{BC}}^{(0)} & \\
	& H_{\text{BC}}^{(0)} 
\end{array} \right] \left[ \begin{array}{c}
({\bf z}^{\mtiny{(1)}}_1)^T \\
({\bf z}^{\mtiny{(2)}}_1)^T 
\end{array}\right] & = &  {\bf 0}
\eea 
where $H_{\text{BC}}^{(0)}$ is the parity check matrix of the code ${\cal C}_{\text{BC}}[\lambda,\lambda,\omega,\rho]$. Since ${\bf z}_1 \neq {\bf 0}$, we may assume that ${\bf z}^{\mtiny{(1)}}_1$ is non-zero without loss of generality. Then we have 
\bean
	H_{\text{BC}}^{(0)} 
({\bf z}^{\mtiny{(1)}}_1)^T & = &  {\bf 0},
\eean 
establishing that ${\bf z}^{\mtiny{(1)}}_1 \in {\cal C}_{\text{BC}}[\lambda,\lambda,\omega,\rho]$. By {\em Case 1}, $w_H({\bf z}^{\mtiny{(1)}}_1) \geq \lambda\rho +1$, which implies $w_H({\bf z}_1) \geq \lambda\rho +1$. 
\item The induction step is as follows. By hypothesis, $w_H({\bf z}_a)  \geq \lambda\rho+1$ for any non-zero vector ${\bf z}_a \in  {\cal C}_{\text{BC}}[2^a\lambda,\lambda,\omega,\rho]$ satisfying $f(X;{\bf z}_a) = 0$. Consider a non-zero vector ${\bf z}_{a+1} \in  {\cal C}_{\text{BC}}[2^{a+1}\lambda,\lambda,\omega,\rho]$ satisfying $f(X;{\bf z}_{a+1}) = 0$. Let us write ${\bf z}_{a+1} = [{\bf z}_{a+1,1}, \ {\bf z}_{a+1,2}, \cdots, {\bf z}_{a+1,2^{a+2}\lambda}]$ where ${\bf z}_{a+1,j}$ is a $(1\times \omega)$ vector if $j$ is odd and a $(1 \times \rho)$ vector if $j$ is even. We generate
\begin{equation}  \label{eq:ya}
{\bf y}_a = [{\bf y}_{a,1}, \ {\bf y}_{a,2}, \cdots, {\bf y}_{a,2^{a+1}\lambda}] 
\end{equation} 
where ${\bf y}_{a,j} \in \mathbb{F}_q^{\omega}$ if $j$ is odd, ${\bf y}_{a,j} \in \mathbb{F}_q^{\rho}$ if $j$ is even and
\begin{equation}   \label{eq:yaj}
{\bf y}_{a,j} = {\bf z}_{a+1,j} + {\bf z}_{a+1,2^{a+1}\lambda+j}, \ \ j \in [2^{a+1}\lambda] .
\end{equation}  
We claim that ${\bf y}_a$ is a codeword of ${\cal C}_{\text{BC}}[2^a\lambda,\lambda,\omega,\rho]$. To establish this claim, it is sufficient to check if ${\bf y}_a$ satisfies the parity check constraints of every local code in ${\cal C}_{\text{BC}}[2^a\lambda,\lambda,\omega,\rho]$. For an arbitrary support set $A_i, i \in [2^a\lambda]$ with respect to $T_{[2^a\lambda,\lambda,\omega]}(\rho)$, let $W$ be the parity check matrix of local code associated to $A_i$. Let $A'_i, i\in [2^{a+1}\lambda]$ denote the supports associated to the topology $T_{[2^{a+1}\lambda,\lambda,\omega]}(\rho)$. Then
\bean
W ({\bf y}_a|_{A_i})^T &=& W ({\bf z}_{a+1}|_{A'_i} + {\bf z}_{a+1}|_{A'_{i+2^{a}}})^T \ = \ {\bf 0},
\eean 
since the support $A'_i$ and $A'_{i+2^{a}}$ correspond to the same local code with parity check matrix $W$. Let $m_i(X;{\bf y}_a), i \in [2^a\lambda]$ be message polynomials associated to ${\bf y}_a \in {\cal C}_{\text{BC}}[2^a\lambda,\lambda,\omega,\rho]$. By \eqref{eq:yaj}, it must be that 
\bea \label{eq:mjya}
m_i(X;{\bf y}_a) =  m_i(X;{\bf z}_{a+1}) + m_{i+2^a\lambda}(X;{\bf z}_{a+1}),
\eea 
for $i =1,2,\ldots, 2^a \lambda$. This implies that
\begin{align*}
f(X;{\bf y}_a) & = \sum_{j=0}^{2^a-1} m_{1+j\lambda}(X;{\bf y}_a) \\
& = \sum_{j=0}^{2^{a+1}-1} m_{1+j\lambda}(X;{\bf z}_{a+1}) = f(X;{\bf z}_{a+1}) = 0,
\end{align*}
since $f(X;{\bf z}_{a+1}) = 0$ by assumption. If ${\bf y}_a \neq {\bf 0}$, then by hypothesis $w_H({\bf y}_a) \geq \lambda\rho+1$. It follows from \eqref{eq:yaj} that $w_H({\bf z}_{a+1}) \geq \lambda\rho+1$. 

\item Suppose that ${\bf y}_a = {\bf 0}$. Then $m_i(X;{\bf y}_a)$ is zero polynomial for every $i$. Since $\mathbb{F}_q$ is of characteristic $2$, \eqref{eq:yaj} implies that
\begin{equation}    \label{eq:f0yzequal}
{\bf z}_{a+1,j} = {\bf z}_{a+1,j+2^{a+1}\lambda}, \ j = 1,2,\ldots, 2^{a+1}\lambda.
\end{equation} 
Let us define ${\bf z}^{\mtiny{(1)}}_{a+1} = [{\bf z}_{a+1,1} \ {\bf z}_{a+1,2} \cdots {\bf z}_{a+1,2^{a+1}\lambda}]$ and ${\bf z}^{\mtiny{(2)}}_{a+1} = [{\bf z}_{a+1,2^{a+1}\lambda+1} \ {\bf z}_{a+1,2^{a+1}\lambda+2} \ \cdots \ {\bf z}_{a+1,2^{a+2}\lambda}]$. 
Because of \eqref{eq:f0yzequal}, we can replace ${\bf z}_{a+1,j}$ with ${\bf z}_{a+1,j+2^{a+1}\lambda}$ (and vice versa) in every parity check equation involving ${\bf z}_{a+1,j}$. Following the same line of arguments that led to \eqref{eq:z1split} in the the base case, we deduce that ${\bf z}^{\mtiny{(1)}}_{a+1}  \in {\cal C}_{\text{BC}}[2^a\lambda,\lambda,\omega,\rho]$ is non-zero. Either by the inductive hypothesis if $f({\bf z}^{\mtiny{(1)}}_{a+1};X) = {\bf 0}$ or by {\em Case 1} if $f({\bf z}^{\mtiny{(1)}}_{a+1};X) \neq {\bf 0}$, $w_H({\bf z}^{\mtiny{(1)}}_{a+1}) \geq \lambda\rho+1$. Hence $w_H({\bf z}_{a+1})  \geq \lambda\rho +1$, completing the induction.
\een 

\end{document}